\documentclass[usenatbib]{mnras}

\usepackage{newtxtext,newtxmath}

\usepackage[T1]{fontenc}
\usepackage{ae,aecompl}

\usepackage{mathtools}
\usepackage{siunitx}
\usepackage{multirow,tabularx,booktabs}
\usepackage{longtable} 
\usepackage{amsmath}
\usepackage{graphicx}
\usepackage{comment}
\usepackage[normalem]{ulem}
\usepackage{BibDef}
\usepackage{color,xcolor}
\usepackage{hyperref}
\hypersetup{
	colorlinks=true,        
	linkcolor=blue,         
	citecolor=blue,         
}

\def\msun{\,\rm{M_\odot}}

\definecolor{mycolor2}{RGB}{50,100,250}

\definecolor{mycolor}{RGB}{250,50,100}

\definecolor{firebrick}{rgb}{0.7, 0.13, 0.13}

\title[Chasing SMBH mergers with {\it Athena} and LISA]{Chasing Super-Massive Black Hole merging events with {\it Athena} and LISA}
\author[L. Piro, M. Colpi et al.]{
L. Piro$^1$,\thanks{E-mail of the corresponding author: Luigi.piro@inaf.it}
M. Colpi$^{2,3}$,
J. Aird$^{4,5}$,
A. Mangiagli$^{8}$,
A. C. Fabian$^6$,
M. Guainazzi$^7$,
S. Marsat$^8$,
\newauthor
A. Sesana$^2$,
P. McNamara$^7$,
M. Bonetti$^{2,3}$,
E. M. Rossi$^{9}$,
N. R. Tanvir$^5$,
J. G. Baker$^{10}$,
\newauthor
G. Belanger$^{11}$,
T. Dal Canton$^{12}$,
O. Jennrich$^7$,
M. L. Katz$^{13}$,
N. Luetzgendorf$^7$
\\
$^1$INAF, Istituto di Astrofisica e Planetologia Spaziali, via Fosso del Cavaliere 100, I-00133 Rome, Italy \\
$^2$Department of Physics G. Occhialini, University of Milano - Bicocca, Milano, Italy \\
$^3$Istituto Nazionale di Fisica Nucleare (INFN), Milano - Bicocca, Milano, Italy \\
$^4$Institute for Astronomy, University of Edinburgh, Royal Observatory, Edinburgh EH9 3HJ, UK\\
$^5$School of Physics \& Astronomy, University of Leicester, University Road, Leicester LE1 7RJ, UK \\
$^6$Institute of Astronomy, Madingley Road, Cambridge CB3 0HA, UK \\
$^7$ESTEC/ESA, Keplerlaan 1, NL-2201 AZ Noordwijk, the Netherlands \\
$^8$APC, AstroParticule et Cosmologie, Universit\'e Paris, CNRS, Astroparticule et Cosmologie, F-75013 Paris, France \\
$^9$Leiden Observatory, Leiden University, PO Box 9513, NL-2300 RA Leiden, the Netherlands \\
$^{10}$NASA Goddard Space Flight Center, Greenbelt, MD 20771, USA \\
$^{11}$European Space Astronomy Centre - ESA/ESAC, Villanueva de la Ca\~nada, Madrid, Spain \\
$^{12}$Universit\'e Paris-Saclay, CNRS/IN2P3, IJCLab, 91405 Orsay, France \\
$^{13}$ Max-Planck-Institut für Gravitationsphysik, Albert-Einstein-Institut,
Am Mühlenberg 1, 14476 Potsdam-Golm, Germany \\}
\date{Accepted XXX. Received YYY; in original form}

\begin{document}
\label{firstpage}
\pagerange{\pageref{firstpage}--\pageref{lastpage}}
\maketitle

\begin{abstract}
 The European Space Agency is studying two large-class missions bound to operate in the decade of the 30s, and aiming at investigating the most energetic and violent phenomena in the Universe. {\it Athena} is poised to study the physical conditions of baryons locked in large-scale structures from the epoch of their formation, as well as to yield an accurate census of accreting super-massive black holes down to the epoch of reionization; 
 LISA will extend the hunt for Gravitational Wave (GW) events to the hitherto unexplored mHz regime. 
 We discuss in this paper the 
 science that their concurrent operation could yield, and present possible {\it Athena} observational strategies. We focus on 
 Super-Massive (M$\lesssim10^7\msun$) Black Hole Mergers (SMBHMs), potentially accessible to {\it Athena} up to $z\sim2$. The simultaneous measurement of their electro-magnetic (EM) and GW signals may enable unique experiments in the domains of astrophysics, fundamental physics, and cosmography, such as the magneto-hydrodynamics of fluid flows in a rapidly variable space-time, the formation of coronae and jets in Active Galactic Nuclei, and the measurement of the speed of GW, among others. Key to achieve these breakthrough results will be the LISA capability of locating a SMBHM event with an error box comparable to, or better than the field-of-view of the {\it Athena} Wide Field Imager ($\simeq0.4\,$deg$^2$)  and {\it Athena} capability to slew fast to detect the source during the inspiral phase 
 and the post-merger phase.
Together, the two observatories will open in principle the exciting possibility of truly concurrent EM and GW studies of the SMBHMs 
\end{abstract}

\section{Introduction}
\label{introduction}

The science cases of both {\it Athena} \citep{ nandra13} and LISA \citep{LISA17} are outstanding, leading to the missions being selected as the 2nd and 3rd Large class (flagship) missions of the ESA Cosmic Vision Programme. Both missions will observe the most energetic and extreme objects in the universe, the supermassive black holes theorised to be powering the Active Galactic Nuclei (AGN) and the loudest sources of low-frequency gravitational waves (GWs) when in binaries.

LISA will explore the 0.1 to 100 mHz GW frequency interval anticipated to be the richest in variety of sources \citep{LISA17}.
In particular LISA  will detect the GW signal from merging supermassive black holes (SMBHs)   of  $10^4-10^7\msun$ detectable out to redshifts as large as $z\sim 20$  \citep{Colpi19}, the long-duration inspiral of stellar black holes around intermediate-mass and massive black holes \citep{EMRI,Amaro2020}, the early  inspiral of stellar black hole binaries \citep{Sesana2016-LIGO-LISA-BHs} and the nearly monochromatic signal emitted by ultra compact binaries, mostly double white dwarfs,  in the Milky Way and its galaxy-satellites, as individual sources and an unresolved foreground \citep{Nelemans2001,Korol2020,Breivik2020,2022arXiv220306016A}.

{\it Athena} is an X-ray observatory  equipped with a large area telescope and a suite of two instruments  that provide unprecedented sensitivities for wide surveys (Wide Field Imager, WFI; \cite{meidinger19}) and high-spectral resolution (X-ray Integral Field Unit, X-IFU; \cite{barret18}) studies. It aims at  exploring the formation and evolution of the accretion-powered objects in the local to the high-redshift universe, and of the hot gas present in the largest cosmic structures \footnote{ {\it Athena} reached almost the completion of Phase B1. In 2022, ESA 
communicated that its predicted cost-at-completion would significantly 
exceed the resources allocated in the framework of the ESA Science 
Program. {\it Athena} is therefore undergoing a design-to-cost exercise, 
aiming at a new mission design, 
consistent with the established cost cap while preserving 
as much as possible the original science goals and payload configuration.
This paper assumes the nominal 
scientific performance of {\it Athena}, which still constitutes the starting 
point for the re-assessment of the science case. Our aim is to provide a landmark  for future updates on 
{\it multi-messenger} joint observations,
and their connections with the {\it Hot and Energetic Universe} science, the key objective which led to the selection by ESA of {\it Athena}  in 2014, and considered as science pillar for its second large mission. 
The consolidated 
science performance of {\it newAthena}  will be known at the end of its Phase 
A, expected to be completed by 2024.}
A main goal of {\it Athena} is the study AGN on a wide range of X-ray luminosities, from a few $10^{41}$ to $10^{47}$ erg s$^{-1}$ \citep{Reines-Comastri2016}.
 The prospects of detecting the emission from central low luminosity AGN from nearby galaxies are of particular interest.
These galaxies may host either intrinsically dim supermassive black holes above $10^8\msun$, or intrinsically brighter (near-Eddington) lower-mass black holes with masses of $10^4\msun-10^7\msun$,  in the interval that LISA will probe.   Our knowledge of the low-mass end of the black hole mass function is tentative at best \citep{Gallo2019}, despite pioneering investigations \citep{Greene2007} and recent major advances \citep{KormendyHo2013,Reines2019,Greene20,Baldassare2020}.  {\it Athena} and LISA promise to  contribute to shed light into a population, that of light supermassive black holes, largely unexplored.


{\it Athena} aims at reconstructing the accretion history of massive black holes, while LISA should reveal the yet unknown merger history of massive black holes in binaries, i.e. a new population predicted to form during the cosmological assembly of galaxies.
The present science cases of {\it Athena} and LISA are thus complementary and individually outstanding, but as we will show here the
{\it additional} science that the concurrent operation of the two missions can achieve
may provide breakthroughs in scientific areas beyond what each individual mission is
designed for. {\it Multi-messenger} astronomy began with the discovery of the first binary neutron star coalescence on  August 17th, 2017.
The GW event, named GW170817, was discovered by the Advanced LIGO and Virgo detectors \citep{2017PhRvL.119p1101A}, and a short gamma-ray burst named GRB~170817A was observed independently by {\it Fermi} and  {\it INTEGRAL} with a time delay of $\sim 1.7$ seconds \citep{Abbott17gw-gamma}.
An extensive, worldwide observing campaign was then launched across the electromagnetic  spectrum leading to the discovery of a bright optical transient in the nearby galaxy NGC4993, and X-ray and radio emission at the transient's position $\sim$9 and $\sim$16 days, respectively, after the merger \citep{Abbott-multimess,Troja17,Hallinan17}.  This one source  allowed astronomers to confirm
the association of short GRBs to relativistic jets produced by NS mergers,  that can be observed also off-axis from an earth's observer  \cite[e.g.][]{2019Sci...363..968G,Ryan20}, and the production of post-merger neutron-rich ejecta (kilonovae) as sources of heavy $r$-process elements in the Universe \cite[e.g.][]{Metzger20}. Furthermore,
 because of the large distance that GWs travelled from the source to the observer, the joint GW and EM observation of GW170817 led to the first empirical bound on the  propagation speed of GWs, on the mass of the graviton \citep{Abbott-test-GR2019}, and the first measurement of the Hubble constant using GWs as cosmic ladder \citep{Abbott-H0-2017Natur.551...85A}. Extraordinary as they are, these results represent just the first leap toward our exploration of the  GW universe. They further demonstrate the power of joint multi-messenger observations.

Multi-messenger astronomy which combines low-frequency GW observations by LISA with contemporary or follow-up X-ray observations of the same source by {\it Athena} may yield unique tests in the domains of astrophysics, physics and cosmology.  
The possibility of performing these tests depends critically on LISA’s capability to localize the source
with progressively increasing accuracy as the amplitude of the GW signal increases, observe as quickly as possible with {\it Athena} the LISA error box to search for the possible X-ray counterpart of the GW event.
Various classes of sources  can exploit the synergy between  {\it Athena} and LISA \citep{Piro22}.
This paper focuses  on the opportunities opened by {\it simultaneous} observations of SMBHMs
with {\it Athena} and LISA 
as well as on the challenges associated to synergistic observations, with an updated assessment of the localization capabilities of LISA.

The paper is organized as follows: in Sect.~\ref{sec:physical drivers} 
we describe the  astrophysical motivations that justify the synergy between {\it Athena} and LISA. In Sect.~\ref{sec:targets} we present the expected event rates for the classes of astrophysical sources upon which the goals described in Sect.~\ref{sec:physical drivers} can be reached. The paper continues focusing on the most promising family of sources, i.e. merging  massive black holes. Sect.~\ref{MC_section} surveys current theoretical models of the behaviour of matter in the time-variable space-time around SMBHMs.  Sect.~\ref{sect:localization} describes the localization capabilities of the LISA observatory, which leads to the selection of targeted events for {\it Athena}. Sect.~\ref{sect:reference_sky} explores the contribution that the X-ray survey produced by eROSITA \citep{merloni12} may give to the identification of possible counterparts of GW sources with {\it Athena}, discusses possible synergies of {\it Athena} and LISA with IR-optical surveys with particular attention on the Rubin Observatory Legacy Survey of Space and Time (LSST hereon). Sect.~\ref{sect:athena_strategy} describes the {\it Athena} strategies for the identification of the EM counterpart of coalescing massive black holes during the pre- and post-merger phase with the WFI, and post-merger follow-up with the X-IFU. Sect.~\ref{sect:caveats} describes pending uncertainties and discusses the known unknowns.

\section{Physical drivers of GW-EM multi-messenger observations}
\label{sec:physical drivers}

Joint GW and EM multi-messenger observations can answer a number of open, yet unsolved questions related to the nature and origin of GWs, to the environment in which GWs are generated and their propagation properties.  
The additional science that LISA and {\it Athena} can do together touches upon three key domains:

\begin{itemize}
    \item Astrophysics 
    \begin{itemize}
  	    \item Magneto-hydrodynamics of fluid flows in violently changing space-time;
        \item Formation of an X-ray corona and jet launching around newly forming horizons;
        \item  Accretion disc structure.
    \end{itemize}
    \item Fundamental Physics
	\begin{itemize}
	    \item Testing General Relativity as theory of gravity;
	    \item Measuring the speed of GWs and dispersion properties.
	 \end{itemize}
    \item Cosmography
	\begin{itemize}
	    \item Testing the expansion rate of the universe.
	\end{itemize}
\end{itemize}	     
	     
The synergy between LISA and {\it Athena} relies on a number of prospected GW sources.  
These are {\it supermassive black hole mergers} (SMBHMs) in gas-rich environments;  {\it extreme or intermediate mass ratio inspirals} (EMRIs/IMRIs)  where a stellar or an intermediate black hole is skinning the horizon of a large black hole surrounded by an AGN disc \citep{2011PhRvD..84b4032K, 2014PhRvD..89j4059B, 2014MNRAS.441..900M, 2021ApJ...917...43S, Derdzinski21}; and {\it interacting double white dwarf systems} present in large numbers in the Milky Way Galaxy \citep{Kremer2017,Breivik2018,maoz18}. 
The last systems provide information on the relative strength between GW and matter driven torques, and will be studied in a separate paper as these systems are persistent sources in the GW sky and thus can be targeted in different epochs by {\it Athena}.
 Synergies on {\it intermediate mass black hole mergers}  have been studied in \citep{Saini22}

 In this work we focus on {\it supermassive black hole mergers}. In gas-rich galaxies, these sources  will let us explore for the first time the interaction of matter in a highly dynamical space-time. The phase that precedes coalescence is characterized by a profoundly different space-time compared to the post-merger phase and the ability and strategy to detect GW and EM signals in tandem differs in these two phases. In the pre-merger phase, as the SMBHs spiral in, X-ray emission is expected to be modulated in time with characteristic frequencies correlating with the binary orbital motion and the fluid patterns rising in the circumbinary disc and cavity or gas cloud surrounding the two black holes 
\citep{Farris14,Tang17,Bowen17,Bowen18,tang18,dAscoli18,Gutierrez2022,Cattorini2022}.
A periodic EM signal during the GW chirp would allow measurements of the speed of gravitational waves relative to the speed of light with fractional error as small as $10^{-17}$ \citep{Haiman17}.  
In the post-merger phase LISA can localize the source down to fractions of a square degree, as we show in Sect.~\ref{sect:localization}, so that X-ray monitoring of the sky area indicated by LISA has the potential to reveal the presence of a turn-on AGN and/or the luminous consequences of a shock-driven reassessment of the disc to a new spacetime. 
Optical follow-ups would then allow to identify the galaxy and infer the redshift of the source. As GW sources are "standard sirens" in that their signal contains information on the source luminosity distance \citep{1986Natur.323..310S}, it is possible to measure the expansion of the Universe  with no need of a distance ladder using EM observations to secure the redshift.The counterpart makes the GW source a 'bright standard candle'. The Hubble parameter can thus be inferred out to the redshift at which the EM counterpart can be detected and identified. For $H_0$ the estimated error in its measure ranges from 8\% up to 20\% \citep{Tamanini16,2019JCAP...07..024B}.

\section{Candidates for Athena-LISA  synergy}
\label{sec:targets}

Discovering the EM
counterparts of LISA sources  will be groundbreaking {\it per se}. {\it Athena} likely offers the best opportunity to carry out a dedicated search of a counterpart in the EM domain. 
Two fundamental issues have to be folded in, to appreciate the problem at stake.   
This being an uncharted territory, any {\it prediction} about the EM emission, in particular in the X-rays, and about the rate of SMBHMs with a counterpart relies on theory only, with a rather uncertain and widespread range of predictions.

Providing that the counterpart is indeed a photon-emitter, and that it produces a flux above the instrumental threshold, the challenge is then to {\it identify} the counterpart in a field that will likely count thousands of  sources in the LISA error box.  In this respect the X-ray band, the sensitivity and the field of view catered for by {\it Athena} offer the best combination.

Assuming that the broad-band EM spectrum has an overall shape similar to that observed in SMBHs at the center of active galaxies ($\alpha_{\rm OX}=1.3$, \citet{vasudevan09b}) and a ratio between the radio and X-ray luminosity $\nu L_{\nu}(5{\rm {GHz}})/L(2-10\ \rm{keV}) \lesssim 10^{-4.5}$ for radio quiet AGNs \citep{Terashima03,panessa07}, one can relate the X-ray flux to the optical magnitude or radio flux and then compare the number of field sources expected in the three bands.  For example, the X-ray sky at a flux of $\approx 10^{-15}${} erg~cm$^{-2}$~s$^{-1}$ in the 2-10 keV range is populated with about 3000 sources  per deg$^2$ \citep{Georgakakis08}, while at the corresponding magnitude m$_V\approx 24.3$ and radio flux of $\approx3 \mu$Jy there are about  30 (10) times more contaminating objects in the optical (radio) band \citep{Smail95,Vernstrom16}. A proper {\it characterization} of the source properties, in particular in the time domain, is thus necessary to pin down the candidate out of the many contaminating sources. In this section we estimate the expected rates  of SMBH mergers detectable by LISA and   
that can enable {\it quasi-simultaneous or time-critical observation} with {\it Athena}.

LISA is expected to detect the GW signal from SMBHMs  with total mass  $\approx$10$^{4-7}$~$M_{\odot}$ \citep{LISA17}. Those in the mass interval between a few $10^5\msun$ and a few $10^6\msun$ can be detected out to redshift $z\sim 15.$ The detection rate is highly uncertain, in the range $\approx 10-300$ in 4~years and over the whole redshift range  \citep{2011PhRvD..83d4036S,Bonetti19,2019MNRAS.486.2336D,Barausse2020}. The mass-redshift
distribution  is also subjected to large
uncertainties, and based on modelling, detections
will be dominated in number by lower mass systems at redshift $z>5$, with low Signal-to-Noise ratio ($S/N$). Nonetheless, up to several detections of merging black holes with masses $\ge 3\times 10^5$~M$_{\odot}$ at  $z<2$ are expected per year. These events deliver the highest $S/N$ in GW, with an error box potentially small enough to be observed by {\it Athena}. The GW signal increases with time, from the inspiral phase to the merger, thus the best localization is derived in the post-merger phase, with best case localization down to arcminutes as shown in  Sect.~\ref{sect:localization}. 

In this context, we define a {\it gold} binary as a system such that its localization error derived {\it after} the merger is smaller than the WFI field of view ($0.4$~deg$^2$).  These binaries constitute a sample and our preliminary studies indicate that such sample comprises systems  with masses within $3\times 10^5\,\rm M_\odot$ and $10^7 \,\rm M_\odot$ up to $z\approx 2$ (see Sect.~\ref{sect:localization}) and allows {\it Athena} to search for X-ray emission produced in the post-merger phase. 

For the highest $S/N$ and closer events the source can be localized during the  inspiral phase. This would allow {\it Athena} to point {\it before} the merging takes place. We define a {\it platinum} binary as a system such that its  localisation error, determined at least 5~hours before coalescence, is smaller than the {\it Athena}  WFI field of view. The timing is consistent with the {\it Athena} capability of carrying out a target of opportunity (TOO) in 4 hours. The {\it platinum} sample comprises a fraction of binary mergers with mass within $3-10\times 10^5\,\rm M_\odot$ below $z\approx 0.5$, and thus are likely to be rare.  For the {\it platinum} binaries  inspiral and coalescence could be observed with {\it Athena}, including the intriguing perspective to observe in X-rays the merging  event in the act. Notice that {\it platinum} binaries are a subset of the {\it gold} ones. For the best $S/N$ ratio events and sources best oriented in the sky (see Sect.~\ref{sect:athena_strategy}), a follow-up strategy can be devised whereby {\it Athena} starts observing few days before the final binary coalescence. At this time the localisation error of a {\it platinum} binary is $\approx 10$~deg$^2$, an area that can be effectively covered by tiling WFI observations in about 3 days.

Models predict a very wide range of X-ray luminosity, from none to vigorous, with the major question being to which extent gas around the black hole(s) exists and accretes. Assuming that this happens at about the Eddington luminosity, sources will be easily detected by {\it Athena} up to $z\approx2$ (cf. Sect.~\ref{sect:athena_strategy} and Sect.~\ref{sect:caveats}).

\section{Mapping matter in the space-time of merging massive binary black holes}
\label{MC_section}

\subsection{Gas dynamics around inspiraling and coalescing binary black holes}
\label{subsec: gas-dynamics}

The EM emission properties  from supermassive black hole coalescences  are unknown. 
No transient broad-band AGN like emission that could be attributed to the coalescence of a LISA binary has been observed in the variable sky yet, at any wavelength.  Thus, we have to resort on theoretical models 
to infer characteristics of their light curves and spectra during the inspiral and merging phase.  
 
Joint, contemporary observations of the GW and EM signals require the presence of a rich reservoir of gas present during the GW-driven inspiral phase, possibly in the form of a {\it circumbinary disc} surrounding the binary and of {\it mini discs}, which feed the individual black holes \citep[e.g.][]{Gold2014,tang18,Bowen18,Khan18}. 
 
Circumbinary discs have been extensively studied in hydro-dynamical simulations 
when the binary is far from coalescence to explore its gas-assisted secular orbital evolution \citep[e.g.][]{Haiman09,Cuadra2009,Lodato2009,Roedig12,Farris14,Tang17,tang18,Munoz-Lai2019,Moody2019,Duffell2020,Tiede2020}.
At present, there is consensus that the binary carves a cavity in the gas but that accretion is never suppressed. 
But, a remarkable finding is that this type of environment appears to be present even in the relativistic regime when
the binary dynamics is  GW-driven and the circumbinary disc decouples, the viscous time being longer than the GW-induced inspiral timescale. 
3D general relativity magneto-hydrodynamic simulations show that accretion continues all the way to the merger \citep{Gold2014,Farris2015,Khan18}. 
The system evolves into a non-axisymmetric configuration  with the cavity becoming highly lopsided and filled of 
 tenuous, shocked plasma, in part ejected against the disc wall
 where it loses angular momentum to feed the black holes. This leads eventually to the formation of two narrow streams which periodically  convey mass onto the black holes in the form of `mini discs'  extending down to the innermost stable circular orbit  \citep[e.g.][]{Bowen18,Bowen2019}. \citet{Bowen18} found that an $m = 1$ mode over-density, a `lump', forms at the inner edge of the circumbinary disc so that whenever each stream supplying the mini disc comes into phase with the lump this creates a modulation in the accretion flow at the beat frequency between the binary frequency and the lump’s mean orbital frequency \citep{Bowen2019}.

In summary, spiral waves and asymmetries create 
periodicities in the accretion rates that uniquely mark massive binary black holes in the relativistic regime. Since decoupling may occur just when the two black holes enter the LISA band around $\sim 10^{-4}$ Hz, at a distance of $\sim 80$ gravitational radii, the occurrence of periodic gas flows could be revealed combining GW-EM observations.
Periodicities appear to be a generic feature of these systems and may result in distinctive radiation features that could be detected by {\it Athena} in those nearby binaries (the \textit{platinum binaries}) 
for which sky localization allows for the detection of precursor emission 
during the final hundred to tens of cycles prior to coalescence. 

Concerning gas-dynamics during the merger and post-merger phase, simulations  of magnetized circumbinary discs onto non-spinning black hole binaries \citep{Khan18} have shown that collimated and magnetically dominated outflows emerge from the disc funnel independently of the size, extension and mass of the disc model. Incipient jets form and  persist through the very late inspiral, merger and post-merger phases. During merger proper  the magnetization in the funnel grows, and after merger the jet around the new black hole becomes magnetically powered. The region above and below the new black hole is nearly force-free, a prerequisite for the Blandford-Znajek (BZ) mechanism to be at work. Quite interestingly, 
after a few days from the merger, the EM luminosity reaches values comparable to the Eddington luminosity, enabling
follow-up EM observations, after the GW source has been localized with the highest accuracy.  We finally note that the emergence of jets is also seen in simulations of both  non-spinning and spinning binary black holes inspiraling and merging in hot magnetized clouds \citep{Giacomazzo12,Kelly17,Cattorini21,Cattorini2022}
which might represent the environment of a dry merger between gas poor galaxies.  

\subsection{Light curves and spectra from coalescing binary black holes}
\label{sect:lcspectra}

There is no general consensus on the electromagnetic spectrum emerging from
a coalescing black hole binary \citep{Roedig14,tang18,dAscoli18}, nor on the amplitude of the modulation of the accretion luminosity which tracks variability in the accretion rate, and on whether the luminosity is declining or rising in the approach to the merger \citep{Cattorini2022,2021ApJ...910L..26P,Combi2022}. The broad-band emission is an uncharted territory, and the field is in its infancy \citep{2022LRR....25....3B}.

\subsubsection {Precursor emission}

The {\it precursor emission,} days to hours prior to coalescence, is expected to come from the 
circumbinary disc, the mini discs around each black hole and the cavity wall
filled of hot gas and accretion streams, each contributing at different wavelengths to a different extent.  
When the accretion rate makes the flow optically
thick, soft  (2 keV) X-ray radiation is dominated by the inner edge of the circumbinary disc and hard radiation (10 keV) by the gas in/near the mini discs \citep{Gutierrez2022}.
 Doppler modulation of the light curve in tandem with the GW chirp could rise in presence of the "light bulb" associated to a mini disc around the secondary black hole \citep{Haiman17}, but this modulation could be erased by the non stationarity of the inflows, driven by pressure gradients at least as much as by internal stresses.  Close to merger the mini disc get thinner as the tidal truncation radius 
shrinks reducing the Hill sphere to the size of the hole's innermost stable circular orbit \citep{2021ApJ...910L..26P}. During the GW chirp, the dimming of the light curve, that could be a distinctive signature of the last few orbits of a SMBHM makes the cross-correlation between the EM and  GW signal difficult to be extracted.

Outside thermalized regions and in case of low accretion rate, inverse Compton scattering for coronal emission around the mini discs produces hard X-ray emission \citep{dAscoli18}.  
Additional X-ray variability may arise from 
refilling/depletion episodes caused by periodic passage of the black holes
near the overdensity feature at the edge of the circumbinary disc.
Also Doppler beaming \citep{Haiman17} and gravitational lensing \citep{2018MNRAS.474.2975D} can modulate
the observed light flux seen by near-plane observers. The emission is in general highly anisotropic, especially
when the binary is seen edge-on, and thus with the lowest GW amplitude.

Finally, EM emission nearly coincident with the merger involves the gas present within the  orbit at the onset of the GW-driven inspiral. In this phase, the orbital evolution may become much faster than the viscous evolution within the mini-discs: gas is pushed onto the primary at a rate much faster than the accretion rate that would be provided by the viscous torque, possibly resulting in a super-Eddington flare at merger \citep[e.g.][]{ArmitageNatarajan02,Chang+10,Lodato2009}. In fact, the peak luminosity of this transient emission is highly dependent on the (uncertain) amount of mass present in the mini-discs \citep{Tazzari2015}. Such an energetic transient may be expected to produce copious X-ray emission, however a thorough investigation of its spectral features --as far as we know --has not yet been accomplished.

\subsubsection{Post-merger emission}

The post-merger EM emission may be the more easily discovered by {\it Athena}, because of the smaller LISA localisation error. It might arise  -- in particular in the X-ray band --  from a newly launched jet and/or 
from dissipation of energy within the former circumbinary disc ultimately originated by changes in the spacetime and/or by resumed accretion around the newly formed massive black hole. These counterparts
may be visible on relative long (and uncertain) timescales where the most "prompt" signal comes from the jet after days to months, while a disc brightening --often called "afterglow" -- would take years. In the following, we briefly review these scenarios.

Spinning black holes are powerful engines of jets \citep{B-Z77}, therefore it may be possible that a jet follows the birth of the massive black hole resulting from the coalescence. \cite{Yuan2021} recently investigated the broadband non-thermal emission produced by the jet while propagating through a wind originated pre-merger from the circumbinary and mini-discs. They found that radiation from radio to gamma-rays rises after a time from merger between $(0.003-1)$ year, for  $10^6\msun$ black holes, time primarily determined by the scale height of the circumbinary disc and the viscosity parameter. For a moderate accretion rate, emission persists at detectable levels for months after the jet launch.

GWs carry away energy by an amount equal to $\sim 5-10\%$ of the 
reduced mass-energy of the binary, corresponding to $\lesssim 10^{59}$ erg, for a $10^6\,\rm M_\odot$ equal mass binary. This mass loss weakens the underlying gravitational potential. Additionally, GWs carry away net linear momentum which leads to  {\it gravitational recoil} of the new black hole \citep{Peres62,Lousto10}. This kick velocity is acquired near the time of formation of the common horizon of the merging black holes, and emerges when the two black holes carry either unequal masses, unequal spins, or a combination of the two. The recoil velocity can range between less than 100 km s$^{-1}$ up to a few thousands km s$^{-1}$ \citep{Baker08}.
Both phenomena highly perturb the former circumbinary disc that through shocks strives to recover an equilibrium configuration by dissipating and radiating the energy in excess. However, they cause very different energy dissipation rates, with that from the disc response to mass loss \citep[investigated for e.g. by][]{BodePhinney07,Neill+09,megevand+09,Rossi10,Corrales+10,RosottiLodato12} being lower by a few orders of magnitude \citep{Rossi10}.
The prospects of a EM post-merger transient from a recoiling massive black hole are instead more favourable.
The complex dynamics within the surrounding disc have been studied by a number of authors with a broad range of methods \citep[e.g.][]{Schnittman2008,Lippai+08,Shields+Bonning08,megevand+09,Zanotti+10}. The lightcurve timescale, peak luminosity and overall shape depends on the black hole mass, extent and direction of the recoil and on the disc properties \citep[see e.g. fig. 22 in][]{Rossi10}. For a $10^{6}\,\rm M_\odot$ black hole surrounded by a disc $\sim 1000$ times lighter that receives a nearly in-plane kick of $ \simeq 1000$ km s$^{-1}$, there may be an EM transient rising a year after the merger proper, that reaches a fraction of the Eddington luminosity. However, dedicated calculations of the spectrum of the emission are still missing. We may however anticipate that the spectral shape is highly dependent on where shocks deposit energy through the vertical extent of the disc, with higher energy emission being favoured if energy deposition occurs in the more tenuous layers of the disc atmosphere.

Finally, after the merger, the circumbinary disc may also release accretion energy. Semi-analytical works, based on the assumption that the merger would happen in an evacuated cavity devoid of accreting gas, investigated what happens when the former circumbinary disc is no longer held back by the binary's torque and it viscously spreads inwards towards the newly formed massive black hole. An associated rebrightening in X-rays is then expected with a luminosity that may be super-Eddington: this is sometimes called "accretion afterglow". The timescale for the bulk of the gas to reach the central remnant is several years for a $10^6\,\rm M_\odot$ black hole, but earlier detection just after the merger might be possible \citep[e.g.][]{Milosavljevic2005,Tanaka2010}. We remark however that this scenario should be revisited in light of the current understanding that accretion onto the black hole binary is not suppressed in the last gravitational wave dominated stage of the inspiral: in fact, it persists --producing X-ray emission -- all the way to merger \citep[e.g.][]{tang18}. It is therefore currently unclear if a post-merger detectable "rebrightening" actually occurs and on what (probably shorter) timescale, when considering this configuration where substantial gas lingers close to the newly formed black hole.

\begin{figure*}
\begin{center}
\includegraphics[width=\textwidth]{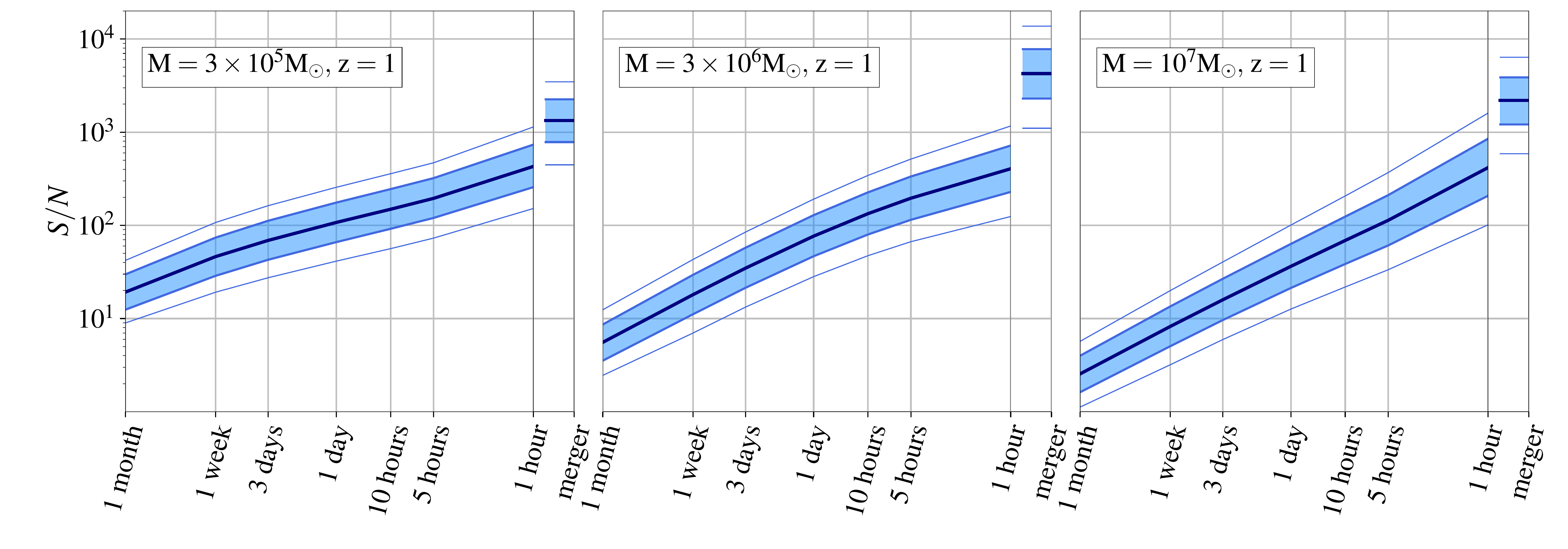}
\caption{  Signal-to-noise ratio ($S/N$) versus time to merger for spinning non-precessing massive black hole binaries with total mass equal to $3\times 10^5\,\rm M_\odot$ (left panel), $3\times 10^6\,\rm M_\odot$ (central panel), $10^7\,\rm M_\odot$ (right panel). The sources are located at $z=1$. The binary mass ratio is extracted random between [0.1-1], the spin between [0-1] and polarization, inclination and sky position angles are extracted randomly from a sphere.   
Shaded areas are the 68\% and 95\% 
confidence interval, computed over $10^4$ systems and the dark solid line 
is the median value.}
\label{SNR-z=1} 
\end{center} 
\end{figure*}

\begin{figure*}
\begin{center}
\includegraphics[width=\textwidth]{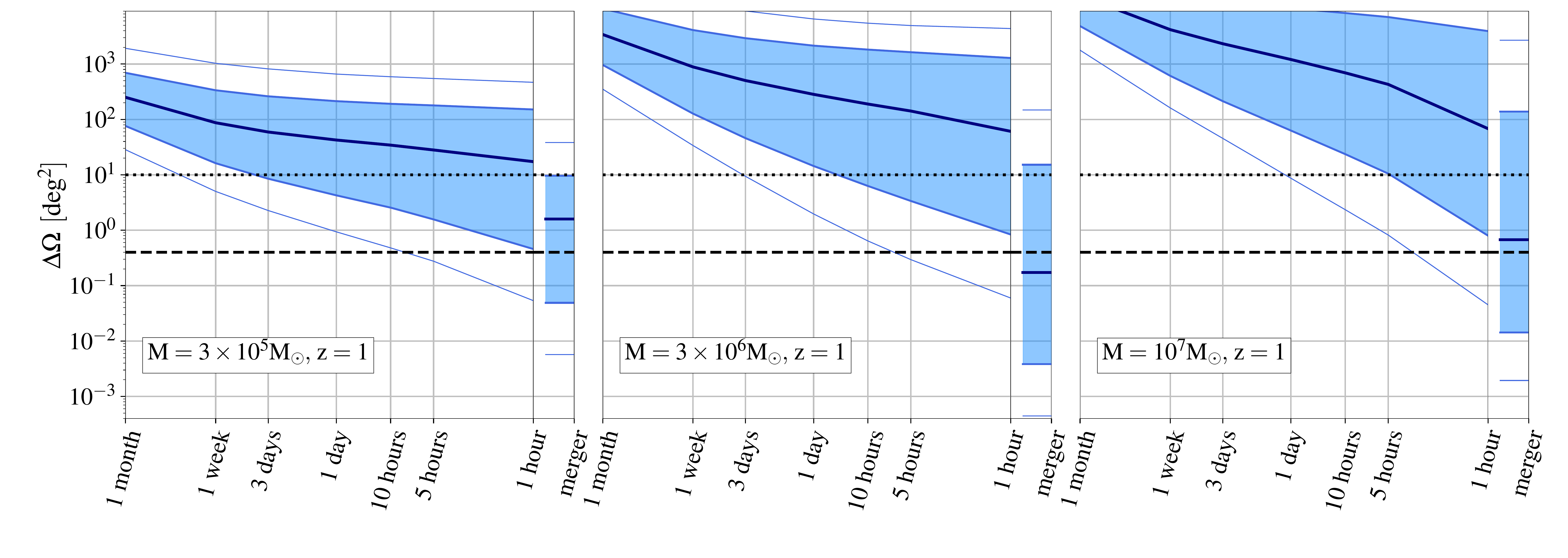}
\caption{ Sky localization error $\Delta \Omega$  (deg$^2$)  versus time to merger for spinning non-precessing binaries as in Fig.~\ref{SNR-z=1}, with total  mass equal to $3\times 10^5\,\rm M_\odot$ (left panel), $3\times 10^6\,\rm M_\odot$ (central panel), $10^7\,\rm M_\odot$ (right panel). The sources are located at $z=1$. Shaded areas are the 68\% and 95\% 
confidence interval computed over $10^4$ systems and the dark solid line 
is the median value.
The horizontal dashed line denotes the field of view ($\sim 0.4$ deg$^2$) of the WFI onboard {\it Athena}. The 10 deg$^2$ wider field of view of LSST  is denoted with dotted line.} 
\label{Sky-z=1} 
\end{center} 
\end{figure*}

\section{Massive Black Hole Coalescences: LISA Sky Localization}
\label{sect:localization}

  LISA is an all sky monitor sensitive to sources at most points on the sky. To build localization information, LISA can exploit two effects. The first is the relatively long duration of the GW signals:  black hole coalescences are observable weeks/days prior to merger, depending on the total mass, mass ratio, orientation of the orbital plane relative to the line of sight and luminosity distance of the binary, so that LISA's orbital motion leaves an imprint on the signal that depends on the position of the source.   The second effect intervenes at merger, when the signal reaches high frequencies: the instrumental response becomes frequency-dependent in a way that informs us
about the signal’s location \citep{Marsat20}. Prerequisite to
achieve best localization is knowledge of the waveform, which
contains information on both the intrinsic and extrinsic parameters of the source, whose estimate is degenerate. Including
higher order modes (HM) associated to the presence of higher-order multipoles in the binary mass distribution has proven to be key in localizing the source \citep{Marsat20,Vishal-Berti2020,Pratten2022}.


In this section, we  focus on a family of  binaries and compute the sky localization error as a function of the time to merger to infer the fraction of those sources behaving as {\it gold} and {\it platinum} binaries.  Later we consider  single-event {\it gold} and  {\it platinum} binaries to discuss sky-multimodality. We then summarize our results on the  localization of the sources  at merger and their detectability with {\it Athena}.

In the process of writing this work, we found larger uncertainties in the sky localization with respect to previous works (in particular with respect to \cite{Mangiagli20} and \cite{Piro22}). Specifically, we found that the sky localization uncertainties for the same system and at the same time before merger might be up to one order of magnitude larger than in the previous results. 
The disagreement between the two approaches resides in:
\begin{itemize}
    \item The previous works adopted an inspiral-only waveform for precessing binaries, while in this study we adopt the spin-aligned model  for the inspiral, merger and ringdwon phases \texttt{PhenomHM}  \citep{London18}. Precession is indeed expected to help breaking degeneracies in the binary parameters, leading to sky-localization uncertainties smaller by a factor $\sim2-5$ \citep{2004PhRvD..70d2001V, 2006PhRvD..74l2001L}. However, from comparisons with previous codes in the spin-aligned case, we found that spin precession alone cannot account for the difference in the results;
    \item  The choice of the reference frame systems where the time cut is applied: it appears\footnote{We could not elucidate this issue entirely, and could only obtain indirect evidence for the origin of the difference.} that in the previous studies the time cut was applied incorrectly in time-of-arrival at the solar systems barycenter, while in this analysis we perform the cut in the time-of-arrival at LISA. As the transformation between the two times depends on the sky position of the source, an incorrect time cut can introduce spurious correlations between the signal termination and the sky position, leading to an artificially optimistic localization. 
\end{itemize}
 The second point especially means that our new code is a better representation of what would happen in reality. Therefore we consider the results in this paper more reliable, albeit more conservative since we are ignoring precession. A more detailed analysis of these differences is left for future investigations.

\subsection{LISA sky localization "on the fly"}
\label{sub:LISAloc}
 In this section, we show results on the LISA localization capabilities "on the fly", i.e. the sky uncertainty $\Delta \Omega$ versus "time to merger", using LISA current design \citep{Robson19,lisa_requirement_docs}. To model the GW signal, we adopt the \texttt{PhenomHM} waveforms for spinning, non-precessing binaries in the inspiral, merger and ringdown phases, which includes the contribution from higher harmonics \citep{London18,Katz20,Marsat20}.  

 We focus on  three representative black hole binary systems with  total mass $3\times 10^5\,\rm M_\odot$ (light), $3\times 10^6 \,\rm M_\odot$ (intermediate), 
and $10^7\,\rm M_\odot$ (heavy), respectively.  We construct a sample varying the binary mass ratios ($q\leq 1$) and spins (aligned with the orbital angular momentum)  extracted from  uniform distributions between [0.1,1], and [0,1] respectively. Polarization, inclination of the angular momentum relative to the line of sight, and sky position angles are uniformly distributed over a sphere. 
Binaries are located at relatively near redshifts $z=0.3$ and $1,$ as a previous investigation by \cite{Mangiagli20} showed that the uncertainties in the sky localization  "on the fly" increase significantly for the sources at larger redshifts.
A Fisher matrix analysis is used, tested on parameter estimations based on the Bayesian analysis of a large sample of binaries (Marsat et al. in preparation). Some limitation of the Fisher-matrix based localization is discussed in the next section.  

In Fig. \ref{SNR-z=1} we show the signal-to-noise ratio $S/N$  as a function of the time-to-merger for the systems located at $z=1$.  Light and intermediate binaries live longer in the LISA band  and accumulate a median $S/N \simeq 10$ already weeks before coalescence,
compared to  heavy systems, which accumulate the same $S/N$  a week or few days before merging.
All these binaries are extremely loud sources  at merger with median $S/N$ in the thousands. We further notice that the dispersion around the median value of both the $S/N$ and $\Delta \Omega$ are  widely spread, as we sampled binaries with varying mass ratios, spins, inclinations and sky positions \citep{Mangiagli20}. 

In Fig. \ref{Sky-z=1} we show the sky localization error, $\Delta\Omega,$ for the same binaries at $z=1$, as a function of the time to merger. None of these sources can be detected by {\it Athena} during their inspiral phase. Only less that 50\% of these sources enter the LSST field of view, days or hours depending on their mass. By contrast, at the time of merger about 50\% of the sources exhibit a localization accuracy consistent with the field of view of the {\it Athena} WFI. It is clear that the best targets are binaries with total mass in the source frame of $\approx 10^6\msun.$ It is remarkable to note that in contrast to the trend of the $S/N$, the spread around the mean for $\Delta \Omega$ is significant, as already found in \citep{Mangiagli20}. 
At the time of merger, sky localization becomes very sensitive to the inclination, polarization angles and actual position of the source in the sky.
In Table \ref{tab:systemerrors} we report the accuracy 
to which the total mass and mass ratio are measured at mergers for the three simulated systems. Given their extreme loudness these parameters are estimated with a very high accuracy and can provide precious information if an EM counterpart is observed.  

\begin{figure*}
\centering
\includegraphics[width=\textwidth]{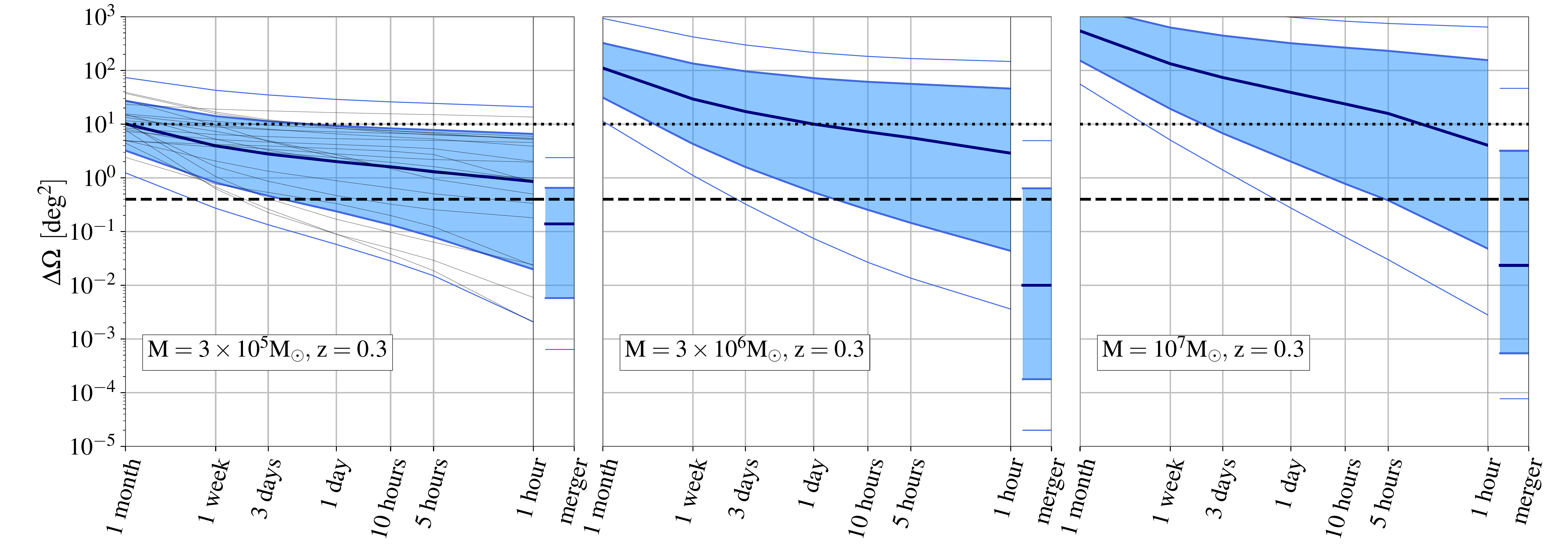}
\caption{ 
Sky localization uncertainty $\Delta \Omega$ (deg$^2$)  for  binaries with total mass in the source frame of $3\times 10^5\,\rm M_\odot$ (left panel), $3\times 10^6\,\rm M_\odot$ (center panel) , $10^7\,\rm M_\odot$ (right panel). The sources are located at  $z=0.3$. Mass ratio, spin moduli and source angles are as in Fig. \ref{Sky-z=1}. Dark red line is the median and the colored areas are as in Fig.  \ref{Sky-z=1}. 
The horizontal  dashed (dotted) line indicates the {\it Athena}/WFI (LSST) field of view. Gray lines in left panel represent a representative sample of the trajectories of mergers for different values of the parameters. }

\label{platinum-mangiagli} 
\end{figure*}

 Fig. \ref{platinum-mangiagli} shows the "on the fly" sky localization uncertainty for sources at $z=0.3$ for the same mass ranges as in Fig.~\ref{Sky-z=1}. The figure includes binaries with parameters as in Fig. \ref{Sky-z=1}. Within this set there are 
{\it platinum binaries}  for which the LISA error box during the inspiraling phase becomes sufficiently small for a follow-up strategy with {\it Athena} to be conceivable. These sources enter the LSST field of view a month before coalescence but less than 50 percent of these  enter the {\it Athena} field of view 10 hours before merger. There are rare cases, for which the number of cycles left before coalescence is around 30, when in the {\it Athena} WFI field of view. This is sufficient for identification of a modulation in the X-ray light curve during the GW chirp \citep{Dal_Canton_2019}. 

\begin{table}
\centering
\caption{Total mass and mass-ratio relative median accuracy at merger for the three simulated systems located at $z=1$.
}
\begin{tabular}{lccc}
& M=3 $\times$ 10$^5$~M$_{\odot}$ & M=3 $\times$ 10$^6$~M$_{\odot}$ & M=10$^7$~M$_{\odot}$ \\ \hline
$\Delta M/M$  & $3.9 \times 10^{-4}$ & $10^{-4}$ & $2.6 \times 10^{-4}$ \\
$\Delta q/q$  & $4.9 \times 10^{-3}$ & $6.6 \times 10^{-4}$ & $1.6 \times 10^{-3}$ \\ \hline
\end{tabular}
\label{tab:systemerrors}
\end{table}
\begin{figure*}
\begin{center}
\includegraphics[width=0.9\textwidth]{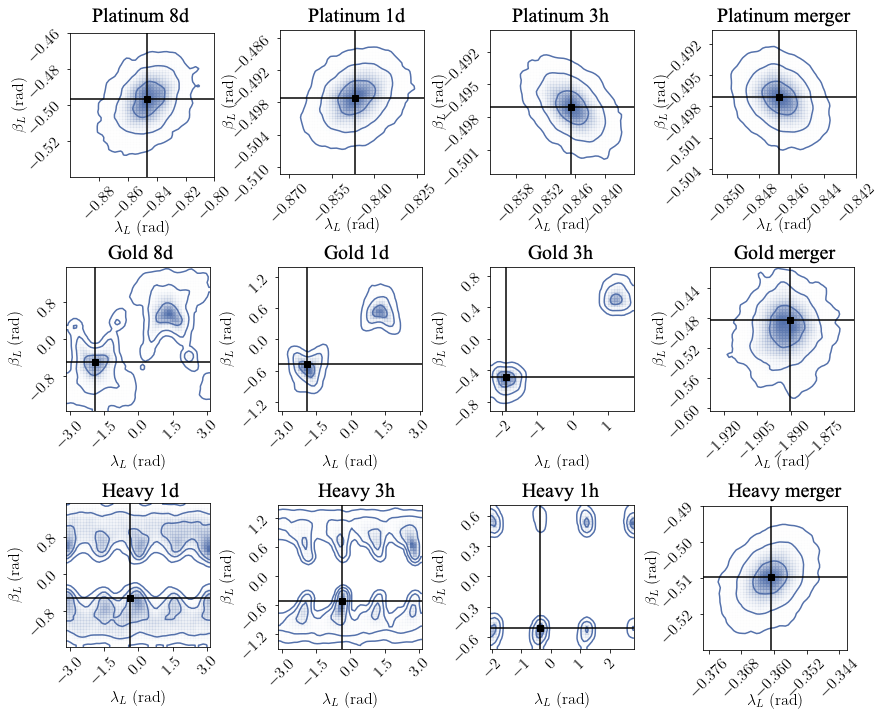}
\caption{Sky--multimodality: Posterior distribution for the latitude $\beta_L$ and longitude $\lambda_L$ (angles in the LISA-frame) for the {\it platinum show-case binary} (top panel) and two {\it gold} binaries (middle and bottom 
panels) with masses $3\times 10^6\msun$  and $3\times 10^7\msun$, mass ratio $q=1/3$, placed at $z=1.$  Posteriors are computed day(s) and hours prior to coalescence, and at merger, as indicated by the lables. Only the {\it platinum} binary shows no multi-modality during the inspiral phase. Solid blue lines correspond to 1$\sigma$, 2$\sigma$ and 3$\sigma$ contours. }
\label{sky-multimod} 
\end{center}
\end{figure*}

 As shown in \cite{Marsat20} frequency-dependent effects in the LISA response at high frequencies and higher harmonics beyond the quadrupole in the GW signal improve localization.  Both are important close to merger and in the post-merger phase. Specifically, higher harmonics break the degeneracy between the inclination angle $\iota$ and the luminosity distance, and the phase-polarization degeneracy. High-frequency effects in the LISA response function allow discrimination between degenerate sky positions, in particular between the antipodal and the true sky position (see next section).
Notice further that for low masses, at low frequencies, most of the information about localization comes from the motion. At merger, high-frequency effects take over and become the dominant source of localization information.
More massive systems have a short-duration signal, and not much information comes from the motion of the detector. Their merger also reaches lower frequencies, so the high-frequency effect is also suppressed and the localization is poorer.

\subsection {Sky map of single-event  {\it Gold} and {\it Platinum} binaries with LISA}
\label{sect:marasat}




 In this subsection we focus on the sky localization error for three single-event binaries, again using the \texttt{PhenomHM} waveforms  which includes higher harmonics during inspiral, merger and ringdown \citep{Marsat20, London18}. The systems we consider here are a {\it platinum}  binary with total mass in the source frame of $3\times 10^5\msun$ at $z=0.3$, and two {\it gold} binaries of mass $3\times 10^6\msun$ at redshift $z=1$, and a heavy binary of $3\times 10^7\msun$ at redshift $z=1$. We take a mass ratio $q=0.3$ and spins aligned with the orbital angular momentum,  with magnitude $\chi_1 = 0.5$, $\chi_2 = 0.2$ for all three systems.
An additional challenge for a secure sky-localisation can be represented by the sky-multimodality, that a Fisher matrix analysis cannot capture. The duration of the signal in the LISA band, and therefore the mass of the system, are crucial for the occurrence of these multimodalities that can  survive post-merger for some systems  \citep{Marsat20}. To explore this, we picked example orientations for our three systems and performed a simulated Bayesian parameter estimation at different times before the coalescence. Contour plots for the resulting sky posterior distributions are shown in Fig. ~\ref{sky-multimod}, in terms of the latitude $\beta_L$ and longitude $\lambda_L$ measured in the frame of LISA, for various time cuts prior to merger. For the {\it gold} binary with {\it intermediate} mass, we can observe that the pre-merger localization shows two well separated maxima corresponding to the source true position and its antipodal point. This is because the antipodal sky position is degenerate for the effect of the LISA motion \citep{Marsat20}. The posterior shrinks as time passes, but the bi-modality remains clearly distinct until coalescence, when the degeneracy is finally removed. The analysis of the {\it platinum} binary (top panels of Fig.~\ref{sky-multimod}) gives a unimodal pattern during both inspiral and merger, as early as 8 days prior to coalescence. This different behaviour is determined by the longer time spent in the LISA band between the detection threshold and the coalescence for this lower mass system. The {\it heavy} system (bottom panels), by contrast, has a very poor localization until 3 hours prior to coalescence. At one hour, the pattern reduces to a characteristic 8-modes pattern \citep{Marsat20}, and the merger finally eliminates all secondary modes. In all three examples, we note that the post-merger localization on the sky following coalescence does not leave ambiguities in its position on the sky\footnote{This is in contrast with the higher-redshfit sources at $z=4$ studied in \cite{Marsat20}, that had a ``reflected'' sky bimodality surviving post-merger.} so {\it Athena} can point at the source precisely including, for the {\it platinum} binary, the inspiral phase as well.

\begin{figure}
\includegraphics[width=0.5\textwidth]{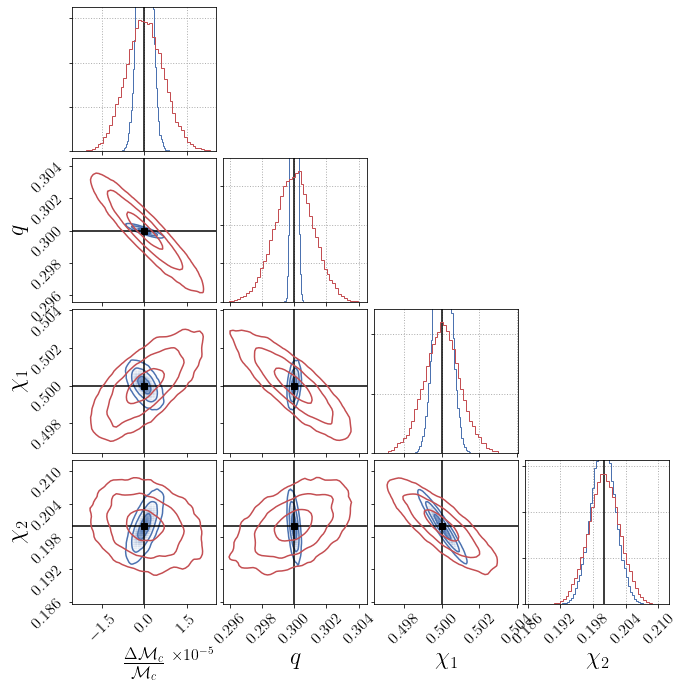}
\caption{Bayesian posterior distribution of intrinsic parameters (chirp mass $\mathcal{M}_c$, mass ratio $q$, and aligned spin components $\chi_1$, $\chi_2$) for an example {\it platinum binary}. The red corresponds to an analysis 3 hours prior to merger, and the blue to a post-merger analysis.  Vertical/horizontal lines indicate the values of the  injection parameters. Contours refer to uncertainties within 1-2-3 $\sigma.$ 
}
\label{plat-corner-intrinsic} 
\end{figure}

 Lastly,  in Fig.~\ref{plat-corner-intrinsic} we show the Bayesian posterior distribution of the intrinsic parameters, i.e.  the chirp mass (defined as ${\cal M}_c=\mu^{3/5}M^{2/5}$ with $\mu$ the reduced mass of the binary), mass ratio and spins, for the {\it platinum} binary, 3 hours before merger and after merger. Although our analysis is restricted to aligned spins, we see that these aligned components are both determined to a few percent accuracy.  

\section{An {\it Athena} observational strategy of SMBH merging events}
\label{sect:athena_strategy} 

\subsection{Fluxes of X-ray counterparts and Athena sensitivity limits}
\label{sect:athena_fluxes}
The above analysis has shown that  SMBH binary coalescences can be localized within the {\it Athena} error box preferentially at merger.
How likely is it that an X-ray source associated to a SMBHM event is sufficiently bright to be detectable
by {\it Athena}? 
In Tab.~\ref{tab:agn} and Tab.~\ref{tab:agnobs} we give the expected fluxes of an AGN hosting a SMBH of mass $M$ (outcome of a merger) and emitting at the Eddington limit, assuming an X-ray to bolometric luminosity ratio of 30 and the
typical spectral shape for an observed AGN (power-law with a photon index $\sim 1.7$).
Tab.~\ref{tab:agn} gives the expected fluxes for an unobscured AGN in the 0.5--2~keV energy band, where {\it Athena}'s sensitivity is highest.
Heavily obscured sources can only be detected at higher energies (2--10~keV, see Tab.~\ref{tab:agnobs} for corresponding flux limits).

\begin{figure}
\begin{center}
\includegraphics[width=\columnwidth]{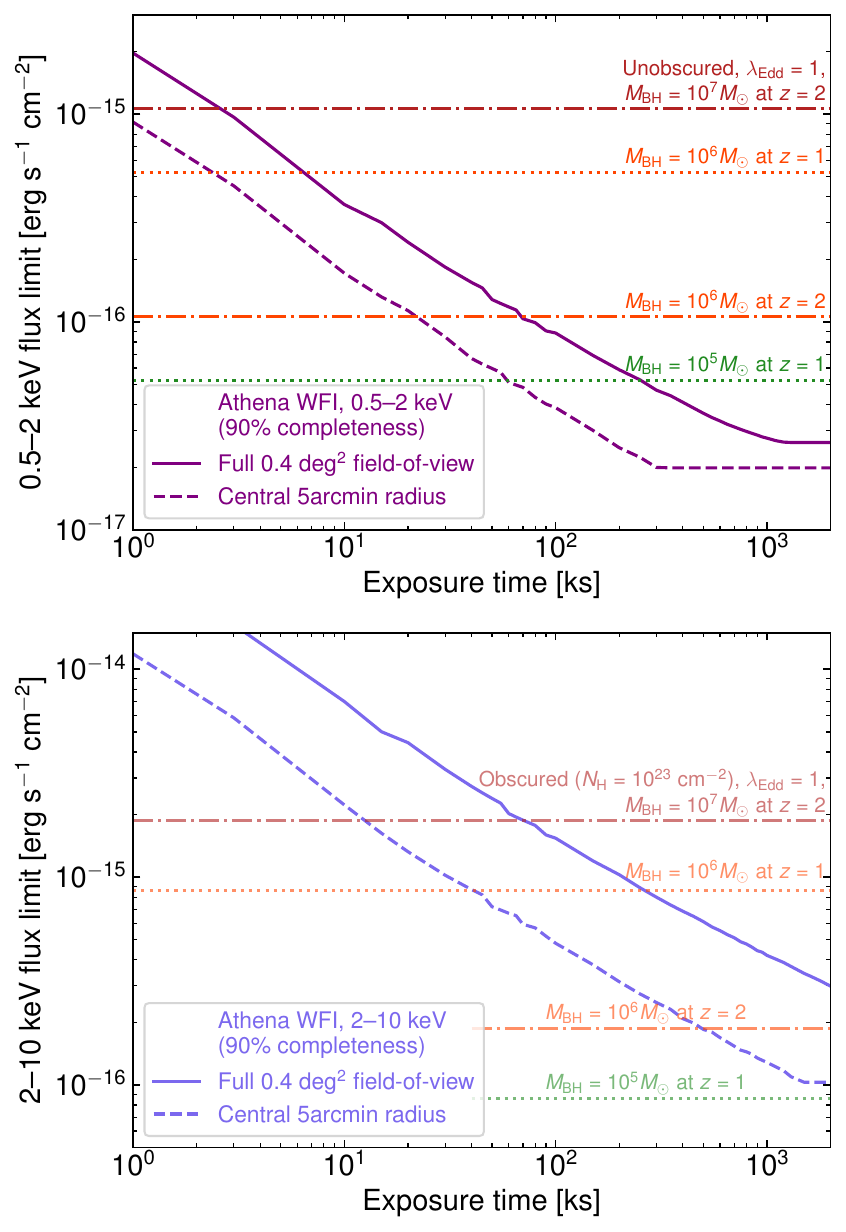}
\caption{{\it Athena} flux limit in the 0.5--2~keV (top) and 2--10~keV (bottom) energy bands as a function of exposure time.
The solid lines indicate the flux limit where {\it Athena} is 90\% complete over the full 0.4~deg$^2$ WFI field-of-view, whereas the dashed line is the flux limit within the central 5~arcmin radius. 
The differences reflect the vignetting and PSF degradation of the telescope at high off-axis angles.
The flattening of the sensitivity curves at the highest exposures is due to the onset of the confusion limit (sources below this limit may still be detected but at a lower guaranteed level of completeness). 
The horizontal lines indicate the fluxes of an unobscured AGN (top) or obscured AGN (bottom) with the indicated masses and redshifts (see Tables~\ref{tab:agn} and \ref{tab:agnobs}).
} 
\label{fig:confusion_limit} 
\end{center} 
\end{figure}

In Fig.~\ref{fig:confusion_limit} we show the flux limits that {\it Athena} is able to reach via observations with the WFI of a given exposure time, in both the 0.5--2~keV and 2--10~keV energy bands.\footnote{Based on the predicted {\it Athena} specifications as of 2019-May-24, see \url{https://www.mpe.mpg.de/ATHENA-WFI/response_matrices.html}.}
The given flux limits are for 90\% completeness over the full 0.4~deg$^2$ FOV (solid curves) or within the central 5~arcmin radius (dashed curves). 
The flux limits are higher over the full FOV as {\it Athena}'s sensitivity drops off-axis due to a combination of vignetting of the telescope and degradation of the PSF. 
The sensitivity is ultimately limited by source confusion, which leads to the flattening of these curves at the highest exposure times. 
To derive the confusion limit, we assume 10~beams per source, with a beam size of radius equal to the half-energy width of the {\it Athena} point spread function (which is $\approx 5''$ on-axis and $\approx 5.9''$ averaged over the FOV), and calculate 90\% completeness limits based on the probability of having a single, unconfused source within the search area.
We note that sources may still be detected below these limits---provided they are not confused with a nearby, brighter sources---but at a lower guaranteed level of completeness (see Sect.~\ref{sect:caveats} for further discussion of the impact of source confusion).

Based on Fig.~\ref{fig:confusion_limit}, we determine that an unobscured AGN associated with a SMBHM of $\sim10^{6}-10^{7}\msun$ at $z>1$ can be detected anywhere within the {\it Athena} FOV in about a few ks, increasing to about 70~ks for lower mass SMBHMs at $z=2$. 
If the associated AGN is obscured (and thus is most efficiently detected at 2--10~keV energies) then the exposure times increase, 
requiring day-long exposures except for the
most massive, and therefore potentially X-ray brightest, SMBHMs.
Lower mass, SMBHMs at $z>2$ that are associated with obscured AGN are likely to remain undetectable, even in extremely deep exposures, due to the impact of source confusion.
The corresponding exposure times are also listed in Tab.~\ref{tab:agn} and  Tab.~\ref{tab:agnobs}.
These numbers provide the rationale for
searching for the X-ray counterpart of a SMBHM event even prior the merging occurs, by optimally scanning a reasonably sized
error box.
\begin{table}
\centering
\caption{0.5--2~keV fluxes (in  erg cm$^{-2}$s$^{-1}$) and exposure times (in brackets) to detect at 5 $\sigma$ a X-ray unobscured AGN at the Eddington limit~with the current configuration of the {\it Athena} mirror+WFI (at 90\% completeness over the full 0.4~deg$^2$ field-of-view).}
\begin{tabular}{lccc}
& M=10$^5$~M$_{\odot}$ & M=10$^6$~M$_{\odot}$ & M=10$^7$~M$_{\odot}$ \\ \hline
$z~=~1$ & 5.3$\times$10$^{-17}$ (250~ks) &5.3$\times$10$^{-16}$ (7~ks) & 5.3$\times$10$^{-15}$ ($<$1~ks) \\
$z~=~2$ &  1.1$\times$10$^{-17}$ ($\gtrsim$1~Ms) &1.1$\times$10$^{-16}$ (70~ks) & 1.1$\times$10$^{-15}$ (3~ks) \\ \hline
\end{tabular}
\label{tab:agn}
\end{table}
\begin{table}
\centering
\caption{As Tab.~\ref{tab:agn} but giving 2--10~keV fluxes for an AGN obscured by a column density $N_\mathrm{H}=10^{23}$~cm$^{-2}$.}
\begin{tabular}{lccc}
& M=10$^5$~M$_{\odot}$ & M=10$^6$~M$_{\odot}$ & M=10$^7$~M$_{\odot}$ \\ \hline
$z~=~1$  & 8.6$\times$10$^{-17}$ ($\gtrsim$1~Ms) & 8.6$\times$10$^{-16}$ (270~ks) & 8.6$\times$10$^{-15}$ (8~ks) \\
$z~=~2$  & 1.9$\times$10$^{-17}$ ($\gtrsim$1~Ms) & 1.9$\times$10$^{-16}$ ($\gtrsim$1~Ms) & 1.9$\times$10$^{-15}$ (70~ks
) \\ \hline
\end{tabular}
\label{tab:agnobs}
\end{table}

\subsection{A WFI follow-up "deterministic" strategy}
\label{sect:athena_deterministic}

\begin{figure}
\begin{center}
\includegraphics[width=\columnwidth]{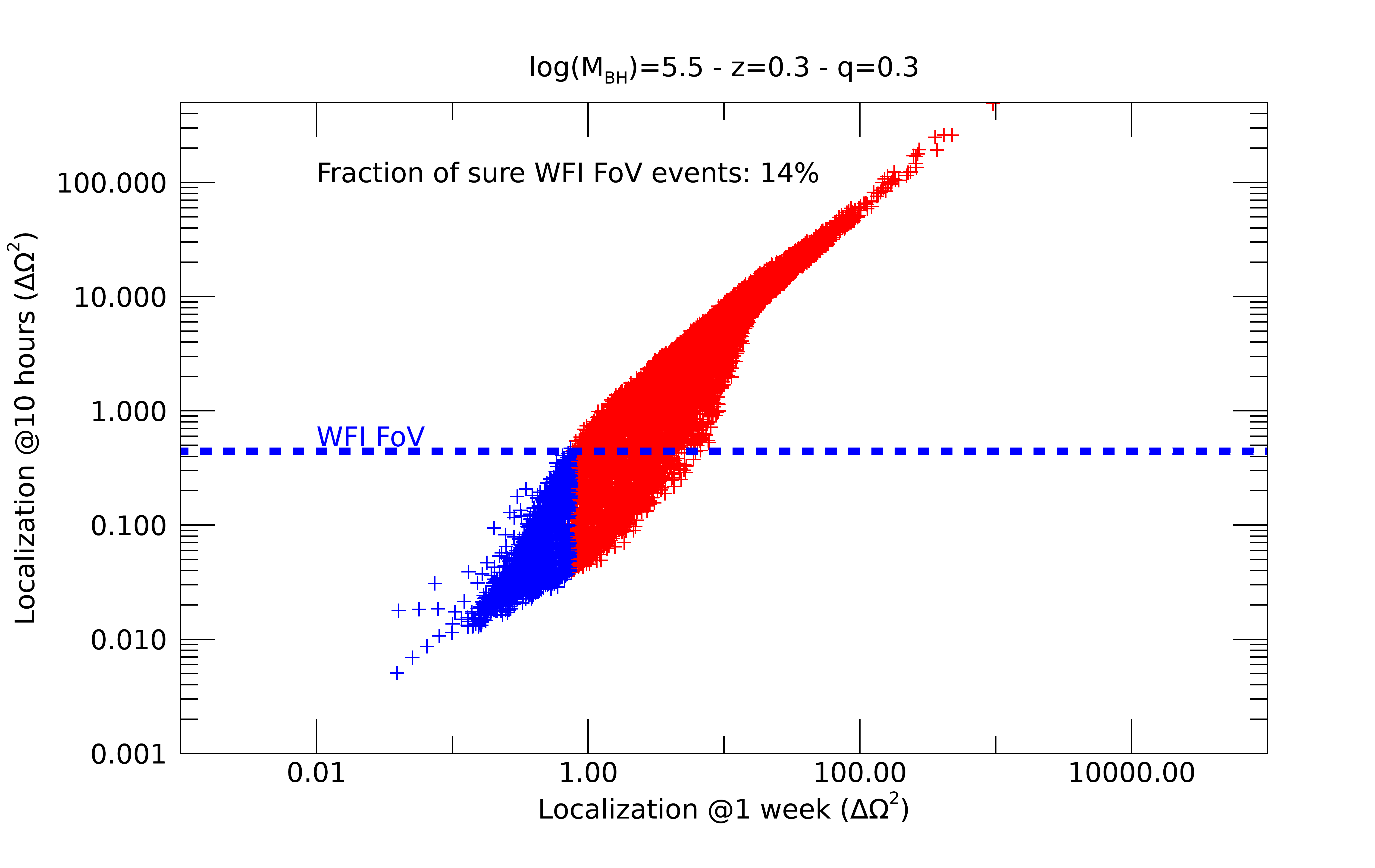}
\includegraphics[width=\columnwidth]{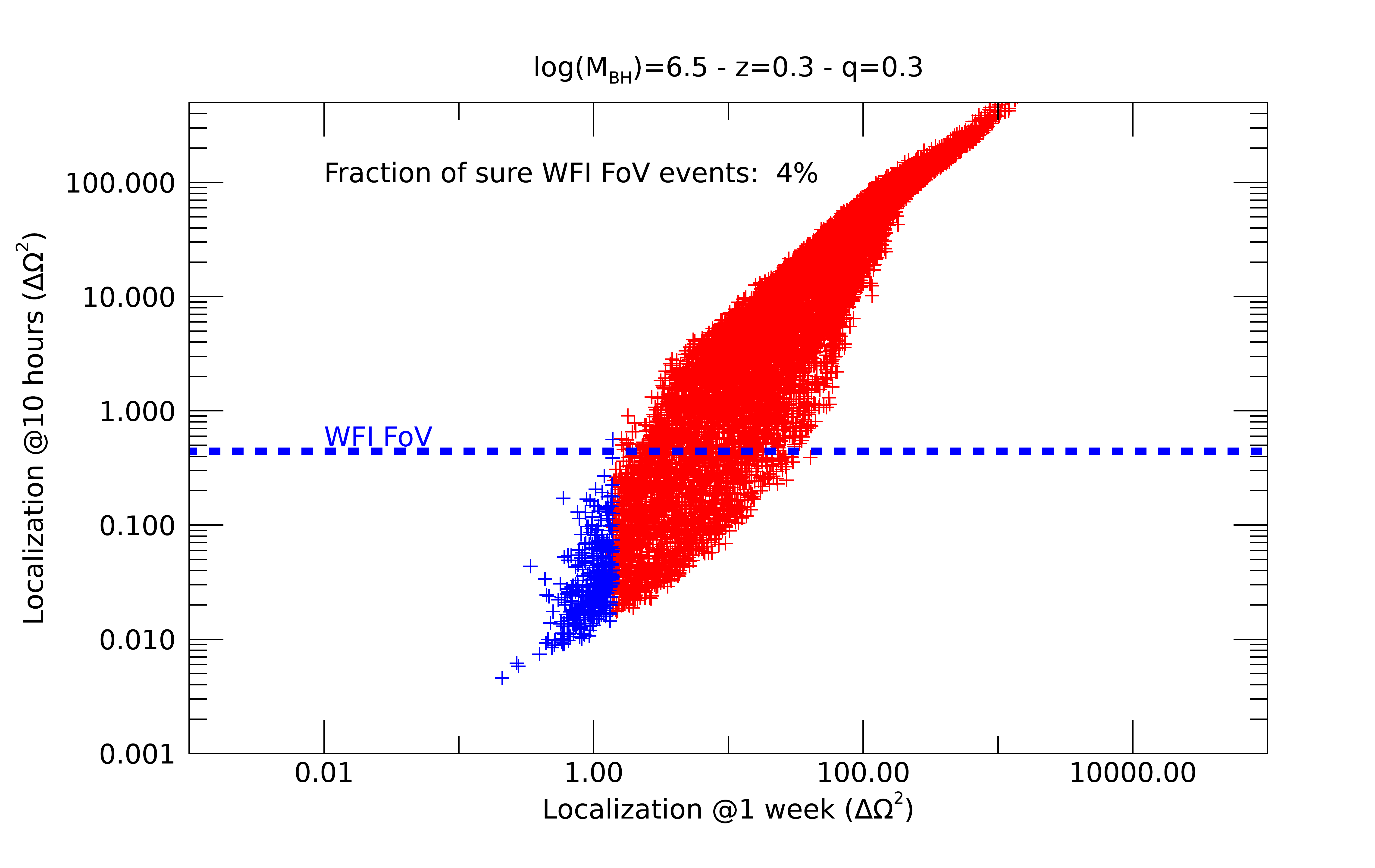}
\caption{LISA localization 10~hours prior to merging as a function of the localization of the same event 1~week prior to merging for $z=0.3$ systems with a total mass
of $3 \times 10^5\msun$ ({\it top panel}) and $3 \times 10^6\msun$ ({\it bottom panel}). The {\it dashed line} indicates the WFI FoV. The {\it inset label}
indicates the fraction of events for which the determination of the localization at 1~week ensures that the localization at 10~hours is better than the WFI FoV
(blue points).} 
\label{fig:deterministic} 
\end{center} 
\end{figure}

The localization of SMBHMs by LISA depends on the cumulative signal-to-noise ratio $S/N$ (cf. Sect.~\ref{sect:localization}) and improves toward coalescence.  While the errors in sky
localization depend on a number of factors (mass, spin, binary inclination, location
in the sky, redshift),
for the best sources (\textit{Platinum} binaries), a localization better then the size of the WFI field-of-view (FoV; $\simeq$0.4~degrees$^2$) is possible a few
hours prior to merging (Fig.~\ref{platinum-mangiagli}). We investigated if the LISA localization evolution may allow future {\it Athena} Project Scientists to establish, with sufficient early warning, if an event will fall within the WFI FoV, such that a re-pointing of the spacecraft is consistent with the operational constraints. For 14\% (4\%) of the $z= 0.3$ events with a mass of $3 \times 10^5\msun$ ($3 \times 10^6\msun $) the localization 1 week prior to merging will ensure {\it deterministically} that the localization accuracy 10 hours prior to merging is smaller than the WFI FoV (the blue data points in Fig.~\ref{fig:deterministic}).
Such events should acquire the highest priority for a follow-up strategy allowing {\it Athena} to observe the sky region of the SMBHM with the WFI.
These fractions vanish to 0 for those $z=0.3$ events with a larger mass, or for events at $z=1$ at all masses.

\subsection{A WFI follow-up "probabilistic" strategy}
\label{sect:athena_probabilistic}

For those events (the majority) on which the strategy described in Sect.~\ref{sect:athena_deterministic} is not applicable, one can still in principle attempt a
WFI follow-up strategy during the inspriraling phase, by performing a raster scan of the most probable, time-evolving LISA localization. This strategy may allow {\it Athena} to acquire a set of X-ray photometric and spectroscopic data points that could unveil a modulation of the X-ray emission while the SMBHM inspiraling proceeds. It is therefore intended to be complementary to the "deterministic" strategy described in Sect.~\ref{sect:athena_deterministic}, which aims at maximizing the probability that the event {\it at merging} is within the WFI FoV, irrespective of prior observations during the inspiral phase.

We hereby require a minimum localization accuracy of 10 deg$^2$ to be achieved 3 days prior to merger in order to trigger this follow-up observational strategy.
An error box of 10~deg$^2$ can
be covered with the {\it Athena} WFI in 3~days with a raster scan of
at least 23 observations of $\simeq 9$ ks ($\sim 2.5$ hours) each. The ``at least'' caveat is primarily driven by the sensitivity of the {\it Athena} telescope decreasing significantly off-axis due
to vignetting and degradation of the point spread function \citep{willingale13}, which depends critically on the ultimate design parameters of the
{\it Athena} optical modules. 
It may be possible to achieve a more homogeneous sensitivity using an overlapping strategy with more pointings with lower individual exposure times.  
With the improvement of the LISA localization the {\it Athena} pointing strategy can be optimized to cover the most likely location of the trigger at any time.
Once the LISA event localization is comparable to, or smaller than the WFI field-of-view,
{\it Athena} could stare to the predicted error box up to the time of the merger. 

In order to estimate the prospective {\it Athena} coverage of SMBHM events triggered by LISA, we run
10000 Monte-Carlo simulations of a possible {\it Athena} observational sequence. We assumed that {\it Athena} starts following the LISA-detected event once the localization is better than 10~deg$^2$.  At each time,
{\it Athena} covers the LISA localization error box with a tile of WFI pointing of equal exposure time - centred to the best-fit LISA position. 
The center of the tile is updated to a new sky position whenever a new estimate of the merging event coordinates is available. When the LISA localization becomes smaller than the WFI field-of-view, we assume that {\it Athena} stares at the best-fit LISA position until a much better localization is available at merging. We assumed also 1~hour overhead time for the transmission and calculation of the LISA coordinates, a 4-hour response time for {\it Athena} to reach the initial position, an {\it Athena} agility of 4~degrees per minute during the raster, and additional 10-minutes of a
"close-loop-slew" at the end of each slew. Once the LISA error box becomes $\le$0.4 squared degrees, {\it Athena} stares at the best-fit error box position
until merging.

\begin{figure*}
\begin{center}
\hbox{
\includegraphics[width=0.65\columnwidth, angle=180]{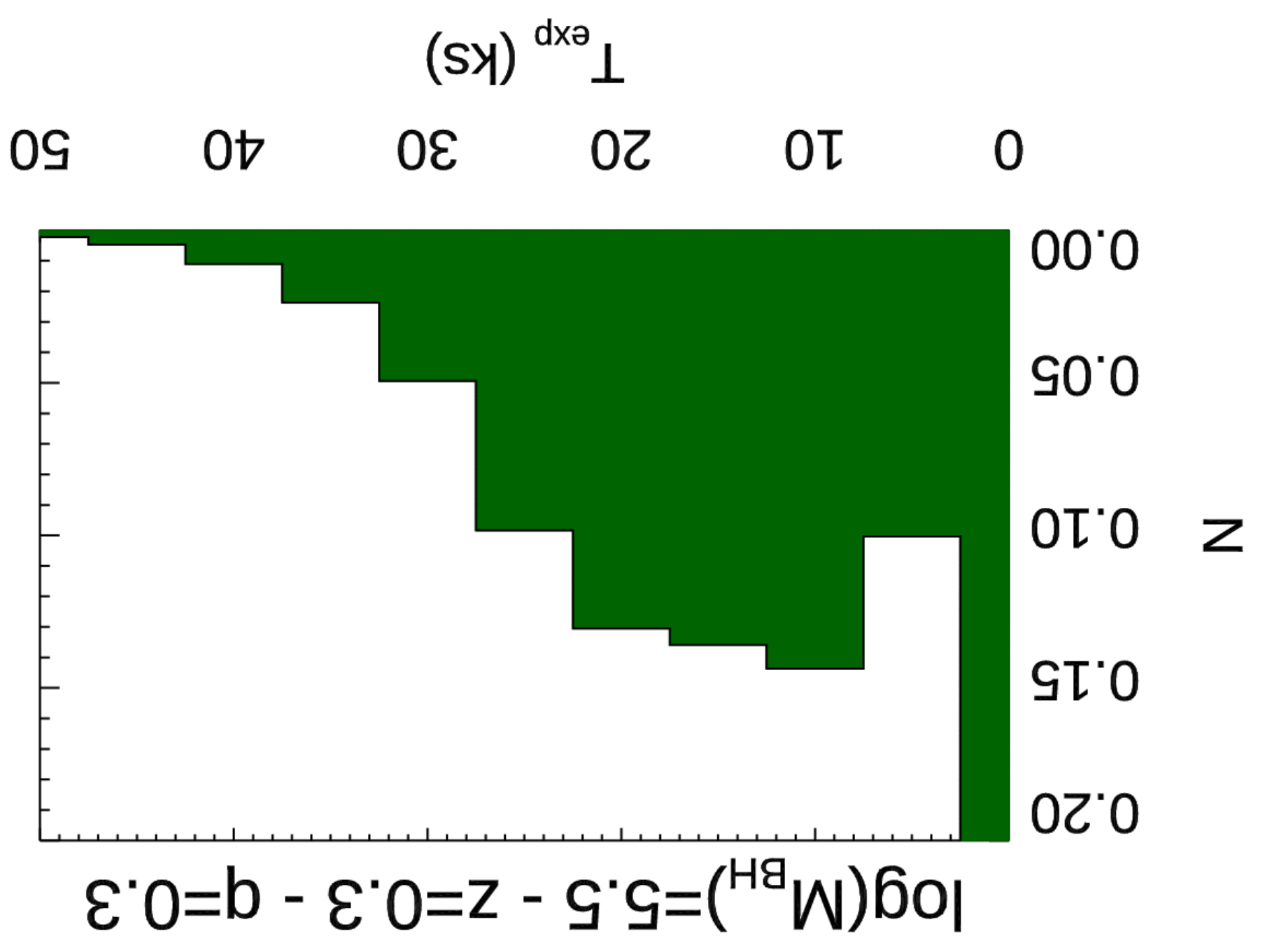}
\includegraphics[width=0.65\columnwidth, angle=180]{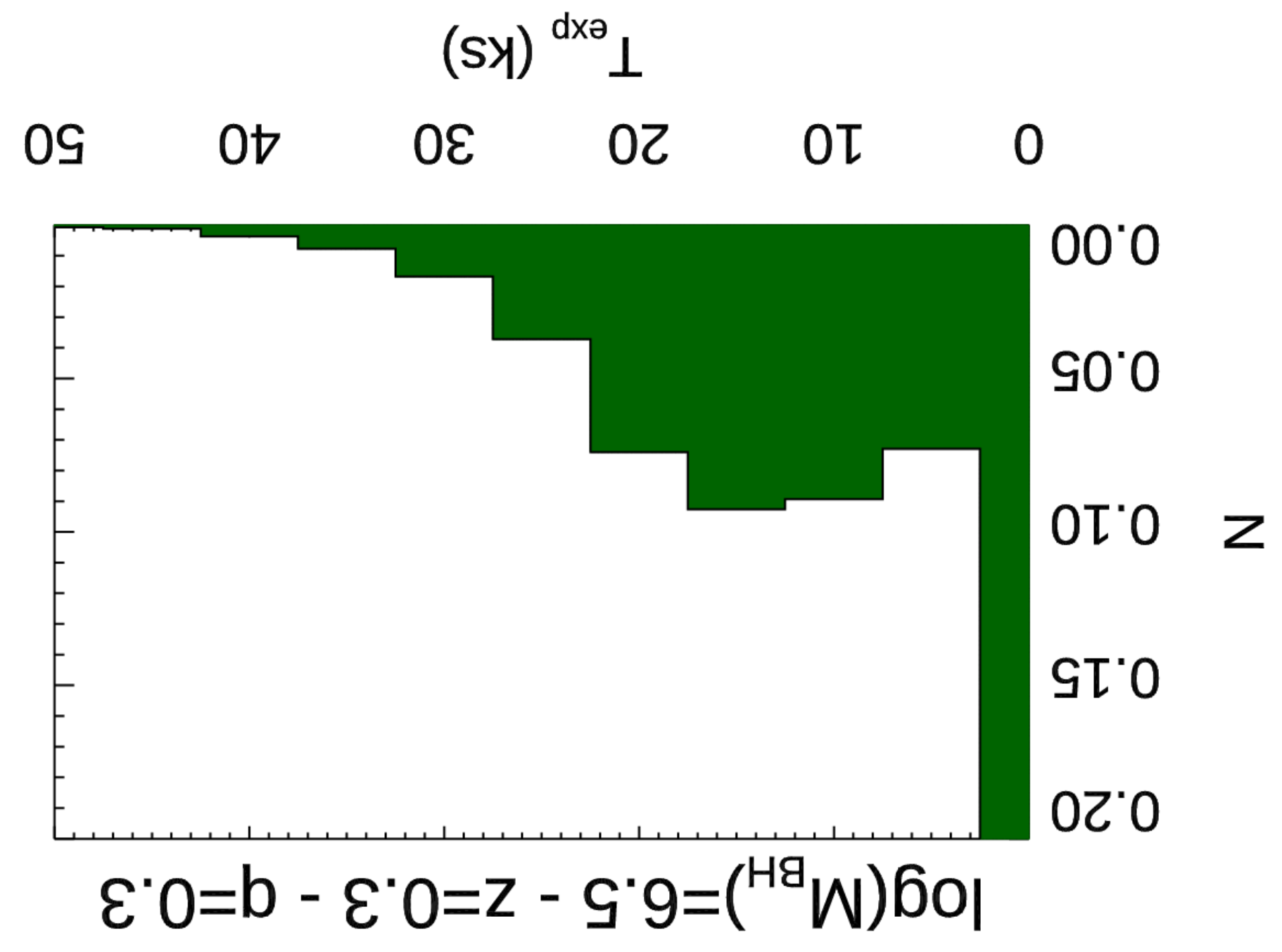}
\includegraphics[width=0.65\columnwidth, angle=180]{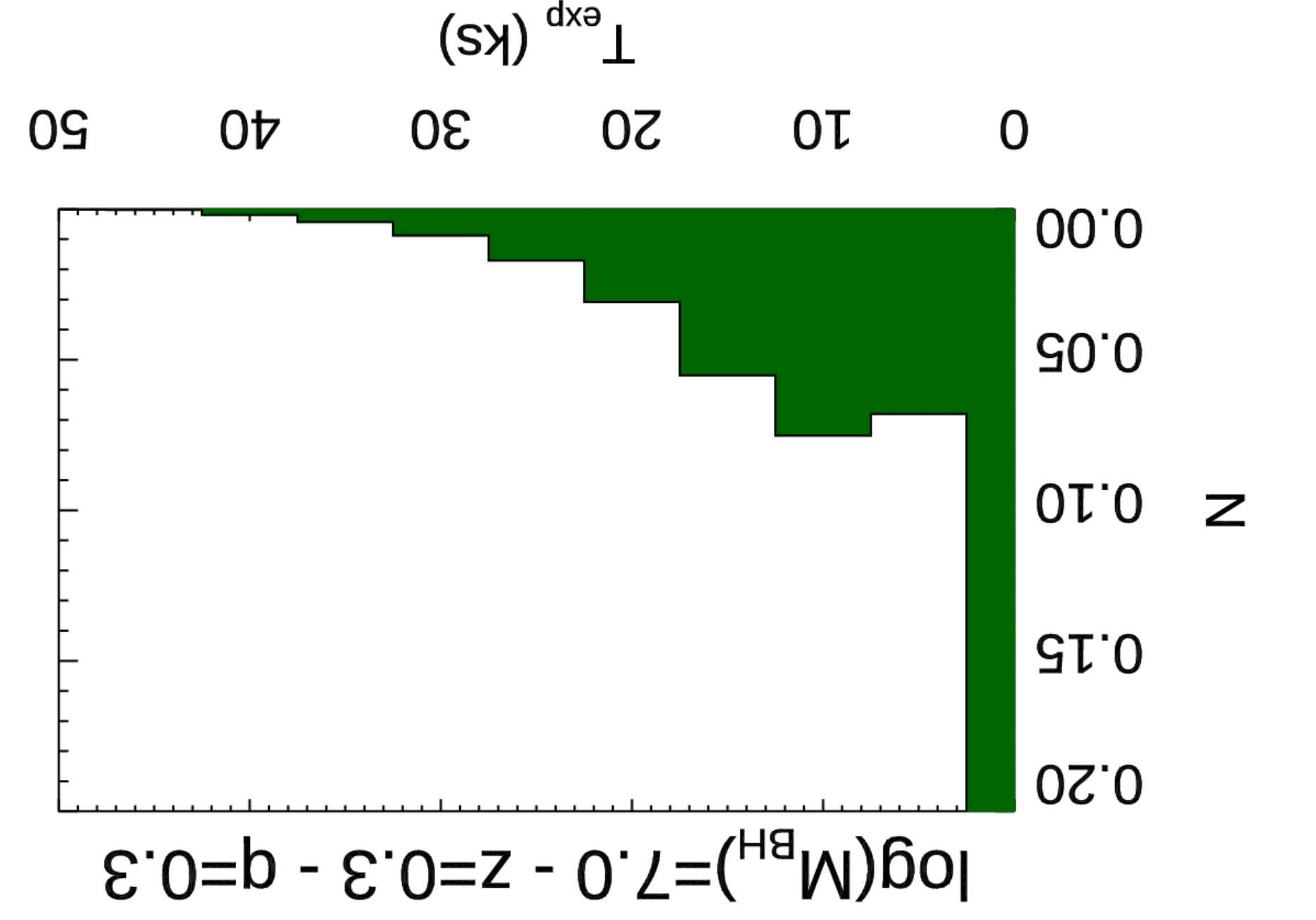}
}
\hbox{
\includegraphics[width=0.65\columnwidth, angle=180]{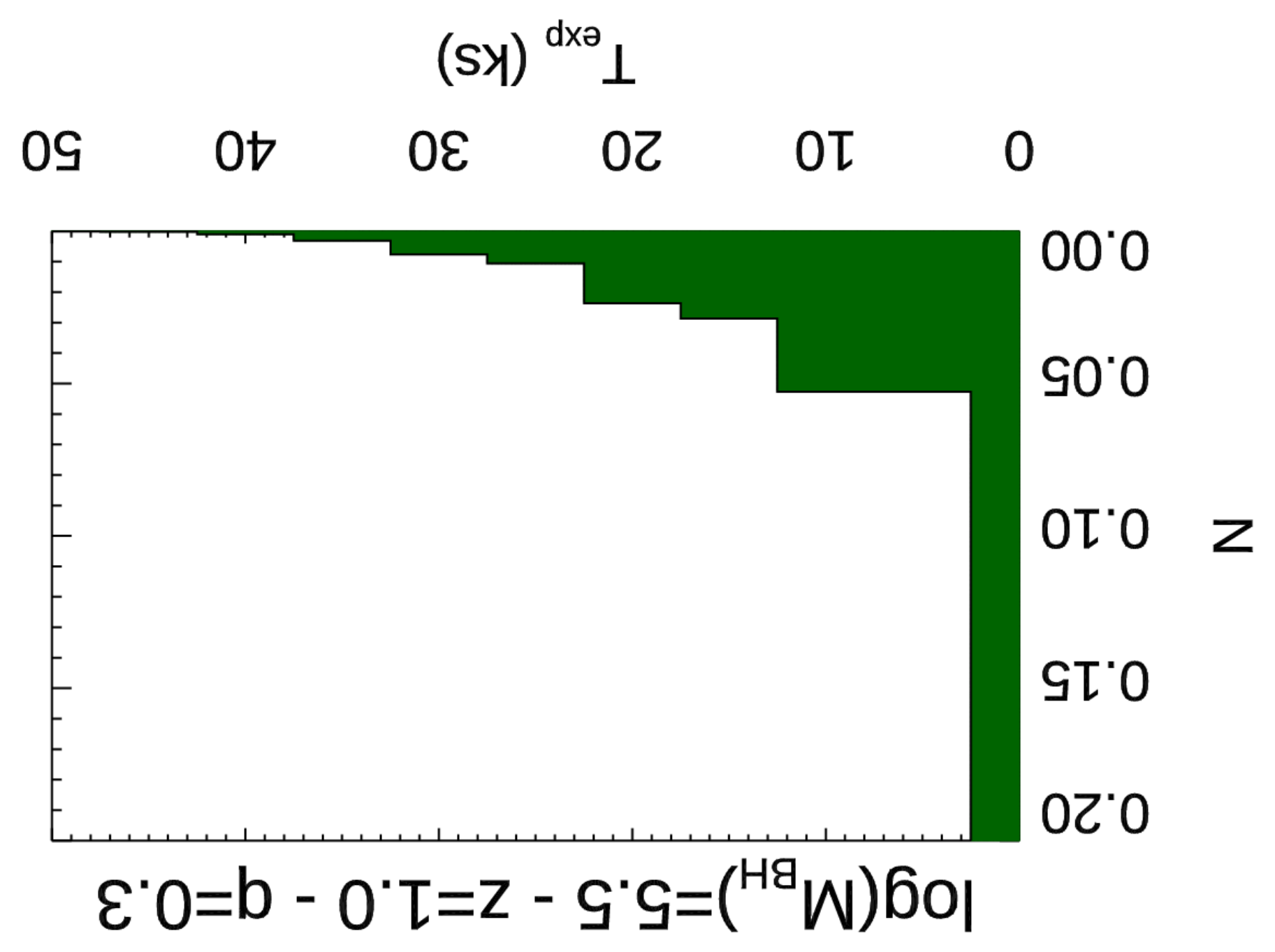}
\includegraphics[width=0.65\columnwidth, angle=180]{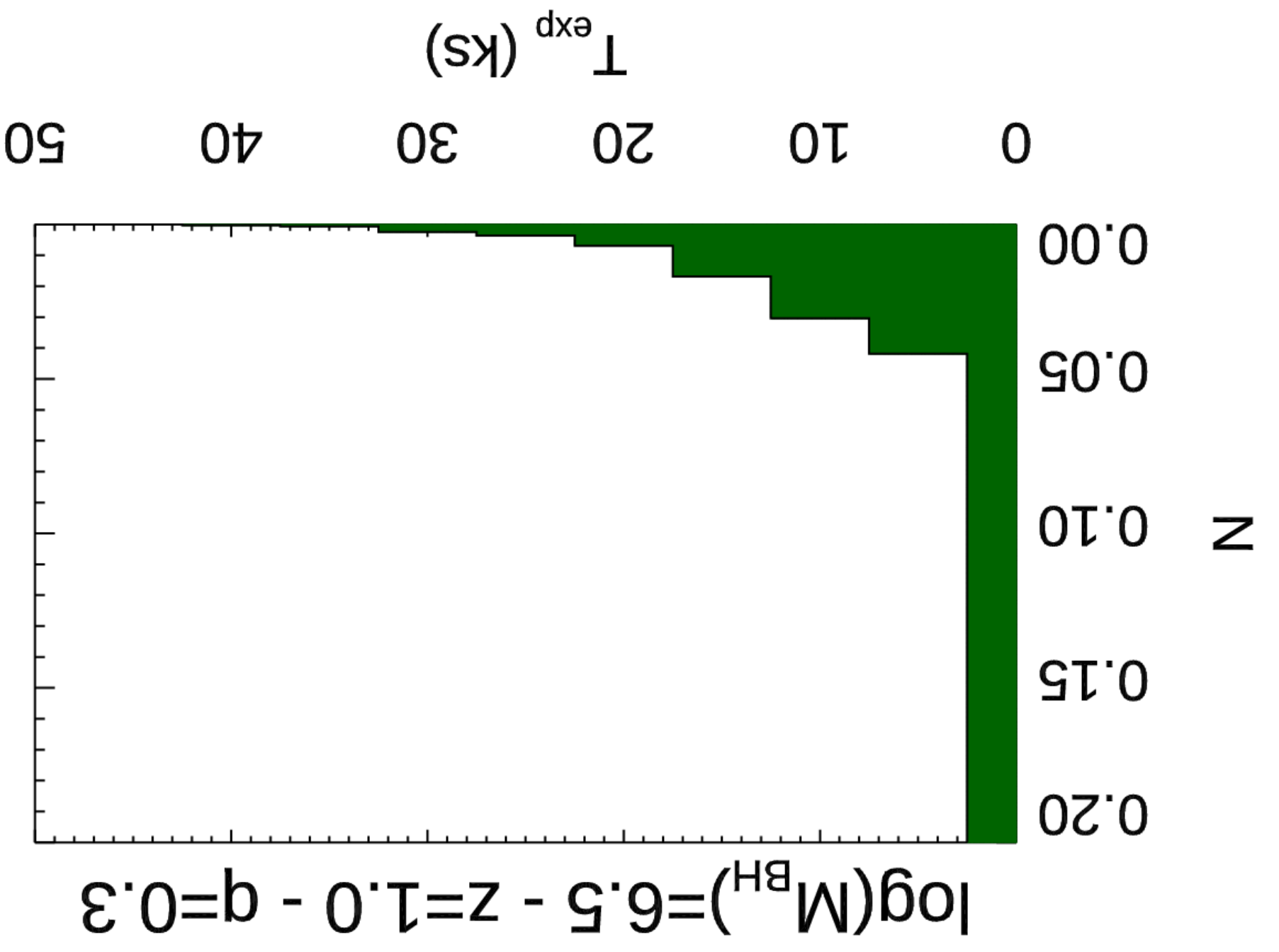}
\includegraphics[width=0.65\columnwidth, angle=180]{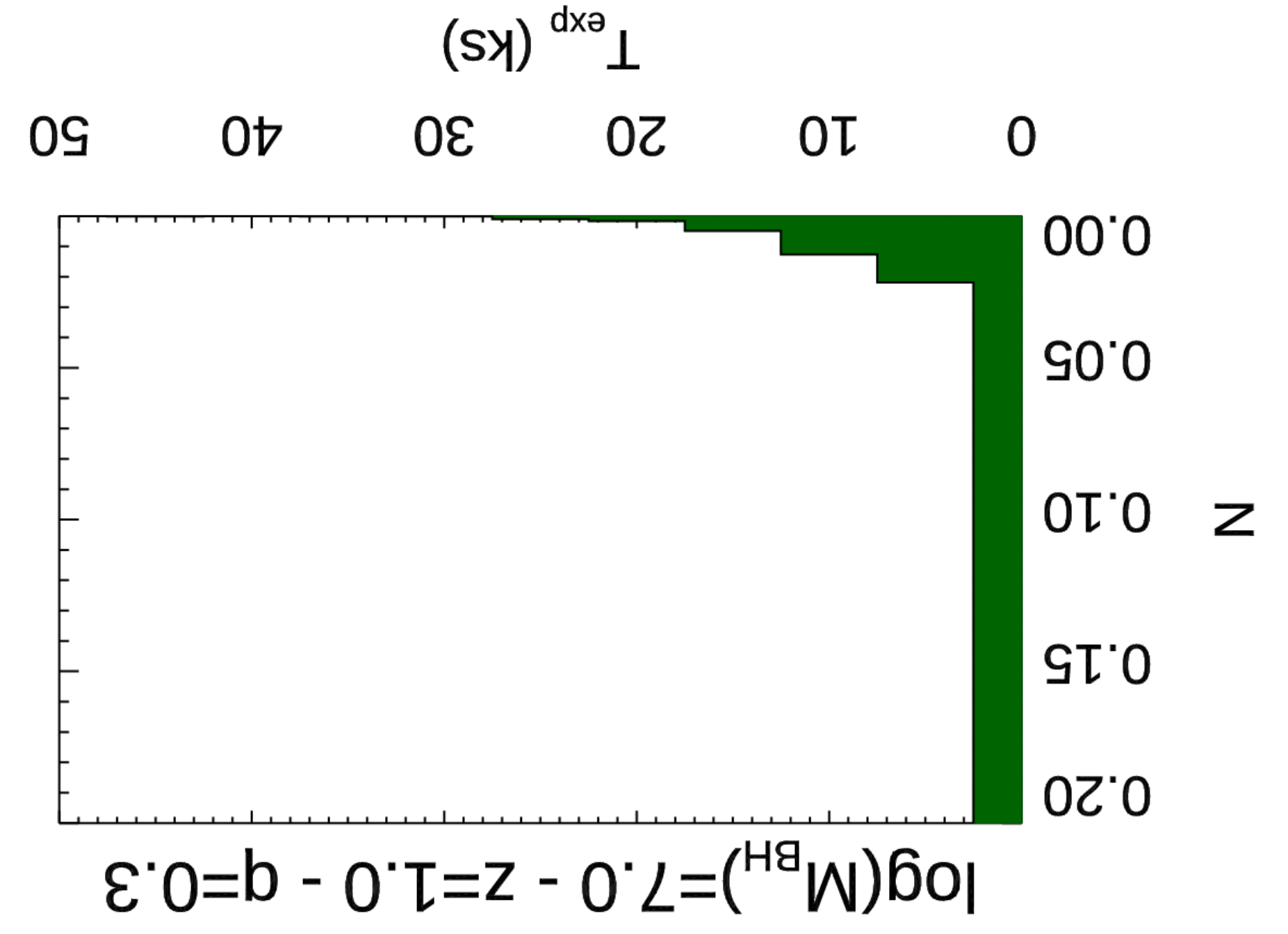}
}
\caption{Distribution of the total exposure during which an event falls within the WFI FoV as a results of the "probabilistic" strategy described in Sect.~\ref{sect:athena_probabilistic}. {\it Top panels}: events with $z=0.3$; {\it bottom panels}: events with $z=1$; {\it left panels}: mass equal to $3 \times 10^5\msun$; {\it central panels}: $3 \times 10^6\msun$; {\it right panels}: $10^7\msun$. The histogram channel corresponding to $T_{\rm exp}$~=~0
extends beyond the upper border of each panel.
}
\label{fig:athena_visits} 
\end{center} 
\end{figure*}

The results of our simulations are summarized in Fig.~\ref{fig:athena_visits}. They correspond to a minimum {\it Athena}
exposure time of 5~ks, after the {\it Athena} spacecraft moves to the next step in the tiling strategy aiming at the LISA error box (unless the LISA localization is
better than the WFI FoV). 
Longer minimum exposure time of, e.g. 10~ks would be $\simeq$10\% less efficient.
The fraction of events that would fall at least once in the WFI FoV ranges between 16\% and 70\% for events at $z=0.3$ (median exposure time $\simeq$10--15~ks), and
between 5\% and 19\% for events at $z=1$ (median exposure time $\simeq$5~ks).
This means that at least for a few events with very large signal-to-noise ratio, the probabilistic strategy is at least conceivable.

Even if a raster scan strategy
of short observations could
allow {\it Athena} to detect the counterpart of the GW-emitting
SMBHM in {\it at least one of the WFI observations}, a significantly more challenging
issue is identifying which of the hundreds of WFI
sources is the true counterpart of the forthcoming merger \citep{Lops2022}. A possible
``smoking gun'' is the variability pattern in the soft and hard
X-ray light curves, mirroring
the GW strain (cf. Sect.~\ref{MC_section}). The expected variability time-scales could vary from minutes to
hours. This implies that it may be hard to disentangle the variability pattern due to
space-time deformation
from the commonly observed variability
in the X-ray light curves of
many classes of celestial sources, most notably AGN, unless at least a few
cycles are observed. The accurate measure of the strain
pattern will represent a key prior in the analysis. Still, this would require a large number of pointed exposures.
An accurate assessment of the minimum combination of visits and exposure time needed to reconstruct
a given energy-dependent variability pattern of the expected light curves is beyond the scope of this paper

In summary, our current understanding of the localization capability of LISA, of its operational constraints, of possible mechanisms producing X-rays in circum-binary discs and mini-discs, of the possible variability pattern of this emission, as well as (and not the least importantly) of AGN astrophysics conspire in making a measurement of the X-ray counterpart of a SMBH binary merging during the pre-merging phase an extremely challenging, albeit exciting, possibility.

\subsection{Super-Massive Black Hole Merger post-merger emission}
\label{sect:postmergerstrategy}

As illustrated in Fig.~\ref{Sky-z=1}-\ref{platinum-mangiagli}, LISA will be able to localize  SMBHMs within the Athena instruments field-of-view when the coalescence has ended. We have shown that SMBHMs can be targeted by {\it Athena}
up to
redshift  $\sim$~2 for 
binaries with masses within $\approx 10^5\msun$,
 and $10^7\msun$. 
This is inferred considering the median of the sky
localization uncertainty. 
A fraction of events between 70\% and 80\% at $z=0.3$, and between 33\% and 56\% at $z=1$  will fall in the {\it Athena} WFI.  (Tab.~\ref{tab:fraction-wfi}). 
 
 Furthermore, 
a {\it gold} binary as well as a {\it platinum} binary has the chance of being localized within an error box as small as 2 arc-minutes, thus  enabling immediate follow-up
with the high-resolution {\it Athena} X-IFU. 
This is possible for a fraction of events between 16\% and 45\% at z~=~0.3, and between 2\% and 19\% at z~=~1 (Tab.~\ref{tab:xifu}). 

\begin{table}
\centering
\caption{Fraction of events whose LISA localization at merger is smaller than the WFI FoV.} 
\begin{tabular}{lccc}
& M=3$\times$10$^5$~M$_{\odot}$ & M=3$\times$10$^6$~M$_{\odot}$ & M=10$^7$~M$_{\odot}$ \\ \hline
$z~=~0.3$ & 74\% & 80\% & 70\% \\
$z~=~1$ &  33\% & 56\% & 46\% \\ \hline
\end{tabular}
\label{tab:fraction-wfi}
\end{table}

\begin{table}
\centering
\caption{Fraction of events whose LISA localization at merger is smaller than the X-IFU FoV.} 
\begin{tabular}{lccc}
& M=3$\times$10$^5$~M$_{\odot}$ & M=3$\times$10$^6$~M$_{\odot}$ & M=10$^7$~M$_{\odot}$ \\ \hline
$z~=~0.3$ & 16\% & 45\% & 38\% \\
$z~=~1$ &  2\% & 19\% & 9\% \\ \hline
\end{tabular}
\label{tab:xifu}
\end{table}
With predictions of tens of events over the mission lifetime (cf. Sect.\ref{sect:caveats}),
several could be followed after the merging occurs to
trace the re-brightening of the disc or the heating of the interstellar medium
by a prompt jet, or a late afterglow due
to gravitational recoil (cf. Sect.~\ref{sect:lcspectra}).
This indicates the truly exciting opportunity to witness
the birth of a AGN. 
A targeted strategy would allow the {\it Athena}
confusion limit ($f_\mathrm{0.5-2keV}\sim$2$\times$10$^{-17}$~erg~cm$^{-2}$~s$^{-1}$, or
$f_\mathrm{2-10keV}\sim$10$^{-16}$~erg~cm$^{-2}$~s$^{-1}$),
within the central 5~arc-min of the WFI field-of-view, to be reached over the LISA error box in $\le$4~days; Fig.~\ref{fig:confusion_limit}).
Monitoring with X-IFU would be possible directly for a fraction of events (Tab.~\ref{tab:xifu}), or {\it post-facto} after the identification of the counterpart
with WFI (at the arcseconds reconstructed positional accuracy).
If {\it Athena} and LISA will be operated simultaneously, a strategy is conceivable whereby a certain numbers of {\it gold} binary fields are monitored periodically post-facto to search for X-ray counterparts, coupled with deep Target of Opportunity observations if/after a counterpart is detected.

\section{The multiwavelength observational context}
\label{sect:reference_sky}

A key priority will be attempting to identify and characterise candidate host galaxies of LISA SMBHMs. Redshift measurements will let us place the LISA event in an astrophysical context.
X-ray observations are likely to be key in this endeavour, since they have the potential advantage of probing merging systems that are less confusion limited.
Recently, \cite{Lops2022} simulated a mock universe to characterize the  X-ray and optical galaxy fields of LISA SMBHMs. For LISA sources of  $ \sim 3\times 10^6\msun$ at $z\lesssim 2$, several tens of AGN emitting in the soft (0.5-2) keV band are present in the galaxy field that can be studied and monitored post-merger. The number of AGN reduces significantly for the  {\it platinum} binaries  of  $3\times 10^5\msun$, leading to an almost unambiguous  identification of the AGN associated to the GW event, i.e. the identification of the X-ray counterpart. Furthermore, by the time {\it Athena} flies a deeper census of the X-ray sky will be available through
the all-sky survey currently being gathered by eROSITA \citep{predehl21}.

\subsection{The reference X-ray sky}

eROSITA will monitor
the full X-ray sky with a sensitivity of $\sim$10$^{-14}$~erg~cm$^{-2}$~s$^{-1}$
in the 0.5-2~keV energy band \citep{merloni12}. 
At the end of the 4-year
survey, about 3 million AGN are expected to be detected up to $z\simeq$3. About one third of the whole
population will be constituted by AGN at $z\simeq$1--2 with a luminosity corresponding to
black holes with masses of $\gtrsim 3\times 10^{7} \msun$
accreting at the Eddington limit (about 10\% thereof are expected to
be X-ray obscured by column densities $>10^{21}$~cm$^{-2}$).
The eROSITA surveys will thus only probe the tip of the AGN population.
However, eROSITA may provide a reference sky template
to refine the selection
of X-ray sources that are {\it unlikely} to be the SMBHM counterparts. \cite{Lops2022} find that from the  eROSITA  survey $\sim 10\%$ of the AGN present in the {\it Athena} field of view at the time of the GW merger event can be discarded, the survey being to shallow to detect the dim AGN associated to a GW event.
It remains unclear what X-ray emission (if any) one should expect from a SMBHM system several
years ($\ge$10) before coalescence. 
Since some simulations predict that the  luminosity  decreases at merger due to the erosion of the mini-discs by the tidal field of the companion SMBH  \citep{2021ApJ...910L..26P}, LISA sources ten years before coalescing could be brighter than at merger. Hence, whether the eROSITA reference sky
could be used to efficiently identify (or rule out) counterpart candidates remains a possibility.

{\it Athena} as observatory will also carry out dedicated large surveys. 
The nominal WFI survey program, designed to address the core {\it Athena} science objects, is expected to cover a maximum of around 50 deg$^2$ to extremely deep flux limits \citep[$f_\mathrm{0.5-2keV}\sim10^{-17}-10^{-16}$~erg~s$^{-1}$~cm$^{-2}$, see][]{Aird13,Rau16}. 
While the chance of an SMBHM event occurring within this footprint is extremely small, these data would provide a deep reference that could be compared to any subsequent follow-up with {\it Athena} (see Section~\ref{sect:athena_strategy} below).
Furthermore, {\it Athena} will have the capability to perform very large area surveys \citep[$>1000$~deg$^2$, see~][]{Zhang20} that would reach flux limits of $f_\mathrm{0.5-2keV}\lesssim10^{-15}$~erg~s$^{-1}$~cm$^{-2}$ (i.e. an order of magnitude fainter than eROSITA), within a reasonable exposure time, and thus could provide a reference X-ray sky that may already identify accreting $10^{6}-10^{7}\msun$ SMBHs out to $z\sim2$. 

\subsection{The reference optical and near-infrared sky}

By the time of any joint {\it Athena}-LISA operations, the Rubin Observatory Legacy Survey of Space and Time (LSST) 
should have finished its wide area sky survey. This is intended to cover nearly half the sky (18000 deg$^2$) to a point-source depth of ABmag$\sim$27.5 in each of six optical/nIR filters ($ugrizy$), detecting up to $10^6$ galaxies per square degree.  
Thus even without further LSST mapping of error regions, most galaxies of interest in the southern sky should already have useful photometry available. This will allow identification of galaxies with photometric redshifts (available for $\sim2\times10^5$ galaxies) consistent with the LISA distance and SMBH mass estimates.  Good candidates can then be targeted for more intensive monitoring,  
to characterise nuclear activity and most important variability. 

It is useful to consider representative low and high redshift scenarios.
Given typical uncertainties in the determination of the photometric redshift of $\sim0.02$--0.03, it is reasonable to take a redshift range of interest of $\Delta z=0.1$.
For a low redshift event, around $z=0.5$, the density of galaxies  (to an absolute magnitude of $M_i=-16$) is $\sim10^4$\,deg$^{-2}$ over this range \citep{capozzi17}. 
Thus  the number of galaxies of interest should be $\lesssim$ 200--1000  either for {\it platinum} events at $z\lesssim 0.5$ in the pre-merger phase and for  10-20 \% of the binaries below $ 10^7\msun$ at  $z\simeq 1$ in the post merger phase \citep[Fig.2, Tab. 4\&5, see also][]{Lops2022}. 
By contrast, at $z>1$ and in particular for  higher mass SMBHMs ($\sim 10^7\rm M_\odot$) the localisation uncertainty area is larger and fields are crowded by   $\gtrsim 10^4$ preventing any identification of a candidate host to the LISA GW event.  Here morphological information will likely allow selection of those galaxies with sufficiently large bulges to contain such a SMBH \citep[but see][for a characterization of host galaxies of LISA SMBHMs] {Volonteri2020}.

The situation in the northern hemisphere is less clear-cut. Facilities such as the Subaru HyperSuprimeCam has the capability to survey at about 1/3 the rate of LSST and so similar mapping may be available for much of the sky not observed by LSST.
Additionally, EUCLID will have finished its wide field survey of $\sim15000$\,deg$^2$ of the high latitude sky.

\section{Caveats}
\label{sect:caveats}

There are many uncertainties involved in forecasting the number of SMBHM events that could be
detected by {\it Athena}. ``Known uncertainties" are discussed here, based on current observations of AGN.
In order for X-ray to be generated before and/or during a merger of a pair of SMBHs, gas must be present 
 in the immediate surroundings and indeed may be instrumental in bringing the SMBHs to a radius where gravitational radiation is strong enough to cause the pair of black holes to spiral together. 
 

It should first be recognised that the X-ray emission detected from AGN by {\it Athena} is dominated by the X-ray corona, which is generally considered to be magnetically powered by an accretion disc orbiting about the SMBHs. The corona is relatively compact and contains energetic electrons with temperatures of tens to hundreds of keV that Compton upscatter blackbody photons from the accretion disc into a power-law X-ray continuum. The observed fraction of the bolometric accretion power emerging in the 2-10 keV X-ray band (the bolometric fraction $f_{\rm bol}$) ranges from about 10  to 2\% or less as the bolometric power increases to  the Eddington limit \citep{vasudevan09,lusso12}. There is as yet no predictive theory of the corona or $f_{\rm bol}$.  Additional 2-10 keV X-ray emission is seen if the object has jets \citep{blandford18}.  There is no observationally based predictive theory for jet occurrence in AGN; a rough guide is that approximately 10\% of quasars are radio-loud due to jets. 

A complication to observing AGN is obscuration. The flat shape of the X-ray Background spectrum in the 2-10 keV band, which is largely the summed emission from all AGN,  demonstrates that most accretion is obscured. Obscuration can occur in all types of AGN, but  simulations suggest that both obscuration and luminous accretion peak in the final merger stages when the two black holes are separated  by less than 3 kpc \citep{hopkins05}, but still far away from the merging phase occurring on micro-parsec scales. This is borne out by 
observations by \cite{koss18} who find that obscured luminous SMBHs show a significant excess (6/34) of nuclear mergers (i.e. a counterpart within 3 kpc) compared to a matched sample of inactive galaxies (2/176). The obscuration most affects the soft X-rays below 2-5 keV.  Prolonged AGN emission at close to the Eddington limit can blow away most of the obscuring gas \citep{fabian09b,ricci17}. 

Violent accretion events such as Tidal Disruption Events (TDEs) could be an alternative template for accretion in the late stages of a SMBH merger. If so, then coronal emission may be weak or absent (see e.g. \cite{Ricci20}), with most of the accretion power emerging from a quasi-thermal blackbody disc, sometimes with jetted emission. Unless jets are formed, X-radiation  from such objects is mostly confined to the soft X-ray band. 

If we assume that accretion takes place in the late merger phase of a pair of SMBH, so that they appear as AGN, we can use the number densities of observed galaxies and AGN to predict the number of final mergers to be expected within a given interval of time. Concentrating on SMBH binaries with masses of $10^6$ to $10^7$\,M$_\odot$ within redshift $z=2$, we start with the number densities of their host galaxies which will have stellar masses of $\approx 10^{9-10}\msun$. \cite{Ilbert13} gives number densities of $10^{-2}$--$10^{-2.5}$~Mpc$^{-3}$ at $z=1$ and $10^{-1.5}$--$10^{-2.5}$ at $z=2$, respectively for 10$^9$-10$^{10}$\,M$_\odot$ galaxies. The probability $p$ that a galaxy has an  AGN accreting above 1\%  of the Eddington  luminosity ($\lambda ={L_{AGN}}/{L_{Edd}}>0.01$) 
has been estimated from observations by \cite{aird18}, giving $p=0.003$ for $10^6$\,M$_\odot$ and $p=0.01$ for $10^7$\,M$_\odot$ SMBHs.  The intrinsic galaxy merger rate is about $10^{-10}$ yr$^{-1}$  \citep{Oleary21}, corresponding to about one major merger per galaxy since $z=1$ \citep{Man16}.  This means that over a 5 yr period of observation  the rate is $2.5\times10^{-10}$ and $2\times10^{-10}$ at $z=1$ and 2 respectively. The rate for dwarf galaxies might be several times less \citep{Deason14} with an average of 0.2 mergers since $z=1$. The galaxy number densities are per comoving Mpc$^3$ and the comoving volume out to $z=1$ is 157 Gpc$^3$ and out to 2 it is 614 Gpc$^3$. Gathering all these factors together, we predict that, per dex in mass and for an observation period of 5 yr, the number of SMBHs of mass $10^6\msun$  merging is $3\times10^{-3}$ within $z=1$ and within $z=2$ it is $3\times10^{-2}$. For SMBH of mass 10$^7\msun$ the corresponding numbers are $3\times 10^{-3}$  
and $10^{-2}$ detectable mergers per 5 yr interval. For SMBH in the mass range $10^5 - 10^6$\,M$_\odot$, we assume a similar mass density to those in the $10^6 - 10^7$\,M$_\odot$ mass range \citep{Greene20}. 

The above predictions (shown in brackets in Tab.~\ref{tab:rates}) assume that the probabilities of a galaxy having an AGN and of it undergoing a merger are independent. If however we assume that all mergers lead to AGN, we can eliminate $p$, which raises the number to those listed in Tab.~\ref{tab:rates}. These are the maximum predicted values, whether or not there is gas in the nucleus. The bracketed numbers in that Table show the expectation if one of the black holes is already an AGN. 
\begin{table}
\centering
\caption{Observation-based predicted upper limit to the number of SMBH merging events visible by {\it Athena} and LISA over 5~years. This is based on the expected merger rates of host galaxies and assumes all mergers lead to an X-ray bright AGN (i.e. $p=1$). The number for $10^5$~M$_{\odot}$ black holes is very uncertain (see text). The numbers in brackets  assume that one of the black holes is already a luminous AGN ($p=0.003-0.01$).}
\begin{tabular}{lccc}
& M=10$^5$~M$_{\odot}$ & M=10$^6$~M$_{\odot}$ & M=10$^7$~M$_{\odot}$ \\ \hline \hline
$z~=~1$ & 1  & 1 (0.003) & 0.3 (0.003) \\
$z~=~2$ & & 10 (0.03) & 1 (0.01) \\ \hline
\end{tabular}
\label{tab:rates}
\end{table}
Some theoretical predictions based on semi-analytical models are not far from the maximum number estimated above, with 7 to 20 EM counterparts in 4~years of joint observations with {\it Athena} in the soft X-rays, and about 2 in presence of obscuration \citep{Mangiagli2022}.
 
Further issues to be noted include source confusion and intrinsic source variability:

\begin{itemize}

\item Source confusion occurs when searching for faint objects at low fluxes. The rising number of even fainter sources increases the probability of 2 or more sources being present in the same detection pixel. This can lead to false detection and at least causes considerable uncertainty in source fluxes, which become biased upward. For the {\it Athena} WFI with a central $5''$ PSF observing the extragalactic sky, source confusion sets in on average at a flux of $f_{\rm 0.5-2 keV}\approx10^{-17}$ erg s$^{-1}$ cm$^{-2}$ for 90\%  of the FOV (Fig.~\ref{fig:confusion_limit}). It is a few times lower in the central region of the FOV.  Sources at fainter fluxes have a higher probability of flux contamination by a second source. If a source position is precisely known (to a fraction of a source pixel) then it may be possible to go slightly deeper using a centred detection pixel. Moreover if the source has an unusual spectrum, or time signature then that can be used to extract source information at lower fluxes.

\item One way in which SMBH pairs in the final merger stage might be detected  is through flux variability  induced by the trans-relativistic orbits of the black holes about each other causing aberration flux changes on the orbital timescale (or twice that if there are two accretion discs). If the GW signal gives the orbital period and its changes in advance of merger then that signal can be searched for  even in a confused source. Intrinsic flux variability is however enhanced for systems with lower mass black holes \citep{miniutti09,ponti12}, making detection of periodic signals more difficult \citep{vaughan16}.   

\end{itemize}

\section{Conclusions}

While the science cases of {\it Athena} and LISA are individually outstanding, the additional science that the concurrent operation of the two missions could achieve may provide breakthroughs in scientific areas beyond what each individual mission is designed for. It encompasses a series of fundamental
questions in modern physics and astrophysics, such as: the dynamics of fluid particles in time-varying, strong
gravity environments; the onset of nuclear activity in the core of galaxies hosting massive black holes; the
physical origin of relativistic jets around spinning black holes, and their launch and interaction with the
galactic environment; the cosmic distance scale; and the measurement of the speed of gravity.

In this paper, we discuss to the possible detection of the X-ray counterparts of coalescing massive black holes in the mass range 10$^5$--10$^7$~$\msun$ that LISA will detect out to large redshifts. Predictions on the detectability of X-ray emission that may rise during the late inspiral and coalescence of the two black holes depend critically on the large uncertainties on the fueling rate and on the hydrodynamical properties of magnetized gas accreting onto the black holes. Within reasonable assumptions, {\it Athena} should be able to detect X-ray emission from sources at $z\le2$. During the inspiral phase (i.e., prior to the merger) X-ray emission could be produced over a wide X-ray
spectral band as thermal (soft) emission from the inner rim of the circumbinary disc surrounding the binary
and/or as coronal (hard) emission from each of the black hole mini-discs within the cavity evacuated by the
spiraling black holes, as well as by shock-heated gas at the wall of the cavity. The X-ray emission could be
modulated with frequencies commensurate with those of the fluid patterns and of the gravitational chirp,
providing the "smoking gun" to identify the X-ray source through a characteristic variability pattern. This
gives in principle the exciting possibility of directly probing, for the first time, the behaviour of matter in
the variable space-time induced by the merging black holes. However, LISA will be able to localize within the field-of-view of the {\it Athena} WFI ($\simeq$0.4~deg$^2$) even the best signal-to-noise events only several hours prior to the merger time. While {\it Athena} will be able to re-point to a random position of the sky pertaining its field of regards within 4~hours, the unambiguous identification of the X-ray counterpart will remain challenging, given the sparse data that even a expensive strategy targeting a wider error box at an earlier time will yield.

The prospective of multi-messenger observations after the merging are potentially more promising. The LISA event error box post-merger could be as small as few
arc-minutes. Pointed observation with the {\it Athena} WFI, or even with the 5-arcminute equivalent diameter X-IFU may allow {\it Athena} to witness the re-birth of an Active Galactic
Nucleus (AGN), or even the launch of a relativistic jet. This will provide a new window for exploring the origin of some of the most powerful and fundamental events in the Universe.

While these unique measurements will undoubtedly represent fundamental breakthroughs in various areas
of physics and astrophysics, one shall bear in mind a series of caveats that make any prediction of the outcome of an actual experiment uncertain. The most significant among them are:

\begin{itemize}
\item the predictions on the nature, and even of the very existence of an X-ray counterpart of a massive black hole merging event detected by LISA are extremely uncertain. However, recent simulations predict
vigorous X-ray emission with a characteristic variability pattern in the pre-merging phase

\item it may be hard to distinguish the modulation pattern in the X-ray light curve induced by the variable space-time around the pair of merging SMBH from the common red noise observed in the field AGN. A quantitative estimate of this effect is beyond the scope of this paper
\item there is, as yet, no observational-based predictive theory of the X-ray corona or relativistic
jets, on which an estimate on the time scale of the formation of an AGN after a massive black hole
merger can be based
\item while there exists no observational data of binary SMBHs with separations lower than a parsec, it is known that dual AGN with separation 1 kpc are typically heavily obscured in X-rays \citep{koss18}. While a sizeable amount of
gas in the environment of the binary black hole is required in order for EM radiation
to be produced, gas can also conspire against the detectability of the X-ray counterparts via heavy
obscuration suppressing the X-ray emission.
\item a concurrent X-ray and GW observations of a SMBHM event requires a fast calculation of the continuously improved GW event locatization by LISA and communication to the {\it Athena} ground segment. In the simulations shown in Sect.~\ref{sect:athena_strategy}, we have assumed 1~hour for the whole process to complete. Such a short time is deemed possible but challenging at our current understanding of the LISA ground segment performance.
\end{itemize}


 EM signs of supermassive black hole binaries (SMBHBs) in the phase anticipating the GW-driven inspiral (which occurs at separation of a few milli-parsecs \citep{Colpi14}) are difficult to discover and disentangle \citep{2022LRR....25....3B}. 
 Signatures of a SMBHB include the presence in the optical spectrum of broad-emission lines Doppler-shifted  relative to the narrow-emission lines, in the case an active SMBH, present in the binary, drags its own broad line region. Alternatively, and at smaller separations, one could reveal periodic  modulation in  the optical or X-ray light curves, which track the orbital motion inside a circumbinary disc, or distinctive X-ray spectral features   \cite[see][for a review]{DeRosa2019}. However,  these features are not unique to accreting SMBHBs as alternative scenarios remain viable  for single AGN \citep{Severgnini2018}.

Significant progress is expected in the coming years, with new facilities from radio to X-rays selecting new potential candidates throughout imaging, spectroscopy and timing, reducing uncertainies in the identifycation of SMBHBs.  In X-rays, systematic searches of light curve modulation and double peaked Fe lines emitted within the mini discs are very promising venues. Future surveys enabled by {\it eRosita}, Einstein Probe and, ultimately, by {\it Athena} should follow-up present SWIFT-BAT results \citep{Serafinelli2020}.  High-resolution X-ray spectroscopy  by cryogenic microcalorimeters on XRISM and then {\it  Athena} will allow detection of fainter double Fe lines, with a separation much below the current limit of $\Delta v/c\approx 3-10 \%$ observed in Si-based detectors \citep{Severgnini2018}, thus enabling detection of  much harder (closer) putative binary systems. 
 For a few SMBHB candidates, the predicted GW signal could be detected by Pulsar Timing Array (PTA) and SKA \citep{Xin2021}. Such a detection would unambiguously demonstrate EM emission from a SMBHB, at least those accessible by PTA coverage, namely    with M $\gtrsim 10^8 \msun$ and separations of the order of hundreds of pc. 

According to this investigation,  the expected number of LISA mergers which are potentially also observable by {\it Athena} based on the observed galaxy density at $z\le2$, and on the observed galaxy merger rates, is $\lesssim$10 over an assumed overlapping operational phase of 4~years \cite[see also][]{Mangiagli2022}. About 20\% of them should correspond to events at $z\le1$, maximizing the probability of X-ray detection by {\it Athena}.


\color{black}{
In summary, the main message of this paper is as follows: 
\begin{itemize}
    \item[-] the multi-messenger concurrent measurement of the GW and X-ray signal from a SMBHM system either in the pre-merger or in the post-merger phase would be a breakthrough result, potentially capable of revolutionizing our understanding of the astrophysics of accreting black holes and fundamental physics alike;
    \item[-] the prospective of measuring in the post-merger phase, when {\it Athena} can stare at a LISA arcminute-level error box, are more promising at our current understanding of the LISA localization capabilities;
    \item[-] due to its unprecedented and innovative nature, any experiment of this novelty is inevitably highly uncertain. This means, at the same time, that the discovery space is potentially vast.
\end{itemize}  
}
\section*{ACKNOWLEDGEMENTS}

This paper is a concurrent effort of the Science Study Teams of the {\it Athena} and LISA missions. The authors gratefully acknowledge advise, suggestions, comments and productive discussions with their members: X. Barcons, D.Barret, K. Danzmann, A. Decourchelle, J.W. den Herder, M. Gehler, M. Hewitson, J. Hjorth, K.Holley-Bockelmann, K.Izumi, P. Jetzer, H. Matsumoto, K. Nandra, J. Nelemans, A. Petiteau, R.M. Sambruna, D. Shoemaker, R. Smith, C.F. Sopuerta, R. Stebbins, H. Tagoshi, J. I. Thorpe, H. Ward, W.J. Weber, R. Willingale. We thank the Director of Science of the European Space Agency, Prof. G. Hasinger, for triggering the process ultimately leading to this paper, and for his continuous encouragement and support.  We thank  an anonymous referee for  useful comments and suggestions. The research leading to these results has been partially supported by the European Union’s Horizon 2020 Programme under the AHEAD2020 project (grant agreement n. 871158). MC acknowledges support by the 2017-NAZ-0418/PER grant. JA acknowledges support from a UKRI Future Leaders Fellowship (grant code: MR/T020989/1). 

\section*{DATA AVAILABILITY}
The data underlying this paper will be shared on reasonable request to the corresponding author.


\bibliographystyle{mnras}
\bibliography{thisbibliography} 

\begin{thebibliography}{}
\makeatletter
\relax
\def\mn@urlcharsother{\let\do\@makeother \do\$\do\&\do\#\do\^\do\_\do\%\do\~}
\def\mn@doi{\begingroup\mn@urlcharsother \@ifnextchar [ {\mn@doi@}
  {\mn@doi@[]}}
\def\mn@doi@[#1]#2{\def\@tempa{#1}\ifx\@tempa\@empty \href
  {http://dx.doi.org/#2} {doi:#2}\else \href {http://dx.doi.org/#2} {#1}\fi
  \endgroup}
\def\mn@eprint#1#2{\mn@eprint@#1:#2::\@nil}
\def\mn@eprint@arXiv#1{\href {http://arxiv.org/abs/#1} {{\tt arXiv:#1}}}
\def\mn@eprint@dblp#1{\href {http://dblp.uni-trier.de/rec/bibtex/#1.xml}
  {dblp:#1}}
\def\mn@eprint@#1:#2:#3:#4\@nil{\def\@tempa {#1}\def\@tempb {#2}\def\@tempc
  {#3}\ifx \@tempc \@empty \let \@tempc \@tempb \let \@tempb \@tempa \fi \ifx
  \@tempb \@empty \def\@tempb {arXiv}\fi \@ifundefined
  {mn@eprint@\@tempb}{\@tempb:\@tempc}{\expandafter \expandafter \csname
  mn@eprint@\@tempb\endcsname \expandafter{\@tempc}}}

\bibitem[\protect\citeauthoryear{{Abbott} et~al.}{{Abbott}
  et~al.}{2017a}]{2017PhRvL.119p1101A}
{Abbott} B.~P.,  et~al., 2017a, \mn@doi [\prl]
  {10.1103/PhysRevLett.119.161101}, \href
  {https://ui.adsabs.harvard.edu/abs/2017PhRvL.119p1101A} {119, 161101}

\bibitem[\protect\citeauthoryear{{Abbott} et~al.}{{Abbott}
  et~al.}{2017b}]{Abbott-H0-2017Natur.551...85A}
{Abbott} B.~P.,  et~al., 2017b, \mn@doi [\nat] {10.1038/nature24471}, \href
  {https://ui.adsabs.harvard.edu/abs/2017Natur.551...85A} {551, 85}

\bibitem[\protect\citeauthoryear{{Abbott} et~al.,}{{Abbott}
  et~al.}{2017c}]{Abbott-multimess}
{Abbott} B.~P.,  et~al., 2017c, \mn@doi [\apjl] {10.3847/2041-8213/aa91c9},
  \href {http://adsabs.harvard.edu/abs/2017ApJ...848L..12A} {848, L12}

\bibitem[\protect\citeauthoryear{{Abbott} et~al.,}{{Abbott}
  et~al.}{2017d}]{Abbott17gw-gamma}
{Abbott} B.~P.,  et~al., 2017d, \mn@doi [\apjl] {10.3847/2041-8213/aa920c},
  \href {http://adsabs.harvard.edu/abs/2017ApJ...848L..13A} {848, L13}

\bibitem[\protect\citeauthoryear{{Abbott} et~al.}{{Abbott}
  et~al.}{2019}]{Abbott-test-GR2019}
{Abbott} B.~P.,  et~al., 2019, \mn@doi [\prl] {10.1103/PhysRevLett.123.011102},
  \href {https://ui.adsabs.harvard.edu/abs/2019PhRvL.123a1102A} {123, 011102}

\bibitem[\protect\citeauthoryear{{Aird} et~al.,}{{Aird} et~al.}{2013}]{Aird13}
{Aird} J.,  et~al., 2013, arXiv e-prints, \href
  {https://ui.adsabs.harvard.edu/abs/2013arXiv1306.2325A} {p. arXiv:1306.2325}

\bibitem[\protect\citeauthoryear{{Aird}, {Coil}  \& {Georgakakis}}{{Aird}
  et~al.}{2018}]{aird18}
{Aird} J.,  {Coil} A.~L.,   {Georgakakis} A.,  2018, \mn@doi [\mnras]
  {10.1093/mnras/stx2700}, \href
  {http://adsabs.harvard.edu/abs/2018MNRAS.474.1225A} {474, 1225}

\bibitem[\protect\citeauthoryear{{Amaro-Seoane}}{{Amaro-Seoane}}{2020}]{Amaro2020}
{Amaro-Seoane} P.,  2020, arXiv e-prints, \href
  {https://ui.adsabs.harvard.edu/abs/2020arXiv201103059A} {p. arXiv:2011.03059}

\bibitem[\protect\citeauthoryear{{Amaro-Seoane} et~al.,}{{Amaro-Seoane}
  et~al.}{2017}]{LISA17}
{Amaro-Seoane} P.,  et~al., 2017, \mn@doi [arXiv e-prints]
  {10.48550/arXiv.1702.00786}, \href
  {https://ui.adsabs.harvard.edu/abs/2017arXiv170200786A} {p. arXiv:1702.00786}

\bibitem[\protect\citeauthoryear{{Amaro-Seoane} et~al.,}{{Amaro-Seoane}
  et~al.}{2022}]{2022arXiv220306016A}
{Amaro-Seoane} P.,  et~al., 2022, arXiv e-prints, \href
  {https://ui.adsabs.harvard.edu/abs/2022arXiv220306016A} {p. arXiv:2203.06016}

\bibitem[\protect\citeauthoryear{{Armitage} \& {Natarajan}}{{Armitage} \&
  {Natarajan}}{2002}]{ArmitageNatarajan02}
{Armitage} P.~J.,  {Natarajan} P.,  2002, \mn@doi [\apjl] {10.1086/339770},
  \href {https://ui.adsabs.harvard.edu/abs/2002ApJ...567L...9A} {567, L9}

\bibitem[\protect\citeauthoryear{Babak et~al.,}{Babak et~al.}{2017}]{EMRI}
Babak S.,  et~al., 2017, \mn@doi [Phys. Rev. D] {10.1103/PhysRevD.95.103012},
  95, 103012

\bibitem[\protect\citeauthoryear{{Baibhav}, {Berti}  \& {Cardoso}}{{Baibhav}
  et~al.}{2020}]{Vishal-Berti2020}
{Baibhav} V.,  {Berti} E.,   {Cardoso} V.,  2020, \mn@doi [\prd]
  {10.1103/PhysRevD.101.084053}, \href
  {https://ui.adsabs.harvard.edu/abs/2020PhRvD.101h4053B} {101, 084053}

\bibitem[\protect\citeauthoryear{{Baker}, {Boggs}, {Centrella}, {Kelly},
  {McWilliams}, {Miller}  \& {van Meter}}{{Baker} et~al.}{2008}]{Baker08}
{Baker} J.~G.,  {Boggs} W.~D.,  {Centrella} J.,  {Kelly} B.~J.,  {McWilliams}
  S.~T.,  {Miller} M.~C.,   {van Meter} J.~R.,  2008, \mn@doi [\apjl]
  {10.1086/590927}, \href {http://adsabs.harvard.edu/abs/2008ApJ...682L..29B}
  {682, L29}

\bibitem[\protect\citeauthoryear{{Baldassare}, {Dickey}, {Geha}  \&
  {Reines}}{{Baldassare} et~al.}{2020}]{Baldassare2020}
{Baldassare} V.~F.,  {Dickey} C.,  {Geha} M.,   {Reines} A.~E.,  2020, \mn@doi
  [\apjl] {10.3847/2041-8213/aba0c1}, \href
  {https://ui.adsabs.harvard.edu/abs/2020ApJ...898L...3B} {898, L3}

\bibitem[\protect\citeauthoryear{{Barausse}, {Cardoso}  \& {Pani}}{{Barausse}
  et~al.}{2014}]{2014PhRvD..89j4059B}
{Barausse} E.,  {Cardoso} V.,   {Pani} P.,  2014, \mn@doi [\prd]
  {10.1103/PhysRevD.89.104059}, \href
  {https://ui.adsabs.harvard.edu/abs/2014PhRvD..89j4059B} {89, 104059}

\bibitem[\protect\citeauthoryear{{Barausse}, {Dvorkin}, {Tremmel}, {Volonteri}
  \& {Bonetti}}{{Barausse} et~al.}{2020}]{Barausse2020}
{Barausse} E.,  {Dvorkin} I.,  {Tremmel} M.,  {Volonteri} M.,   {Bonetti} M.,
  2020, \mn@doi [\apj] {10.3847/1538-4357/abba7f}, \href
  {https://ui.adsabs.harvard.edu/abs/2020ApJ...904...16B} {904, 16}

\bibitem[\protect\citeauthoryear{{Barret} et~al.,}{{Barret}
  et~al.}{2018}]{barret18}
{Barret} D.,  et~al., 2018, in {den Herder} J.-W.~A.,  {Nikzad} S.,
  {Nakazawa} K.,  eds,  Society of Photo-Optical Instrumentation Engineers
  (SPIE) Conference Series Vol. 10699, Space Telescopes and Instrumentation
  2018: Ultraviolet to Gamma Ray. p. 106991G (\mn@eprint {arXiv} {1807.06092}),
  \mn@doi{10.1117/12.2312409}

\bibitem[\protect\citeauthoryear{{Belgacem} et~al.,}{{Belgacem}
  et~al.}{2019}]{2019JCAP...07..024B}
{Belgacem} E.,  et~al., 2019, \mn@doi [\jcap] {10.1088/1475-7516/2019/07/024},
  \href {https://ui.adsabs.harvard.edu/abs/2019JCAP...07..024B} {2019, 024}

\bibitem[\protect\citeauthoryear{{Blandford} \& {Znajek}}{{Blandford} \&
  {Znajek}}{1977}]{B-Z77}
{Blandford} R.~D.,  {Znajek} R.~L.,  1977, \mn@doi [\mnras]
  {10.1093/mnras/179.3.433}, \href
  {https://ui.adsabs.harvard.edu/abs/1977MNRAS.179..433B} {179, 433}

\bibitem[\protect\citeauthoryear{{Blandford}, {Meier}  \&
  {Readhead}}{{Blandford} et~al.}{2018}]{blandford18}
{Blandford} R.,  {Meier} D.,   {Readhead} A.,  2018, arXiv e-prints, \href
  {http://adsabs.harvard.edu/abs/2018arXiv181206025B} {}

\bibitem[\protect\citeauthoryear{{Bode} \& {Phinney}}{{Bode} \&
  {Phinney}}{2007}]{BodePhinney07}
{Bode} N.,  {Phinney} S.,  2007, in APS April Meeting Abstracts. APS Meeting
  Abstracts.
p. S1.010

\bibitem[\protect\citeauthoryear{{Bogdanovi{\'c}}, {Miller}  \&
  {Blecha}}{{Bogdanovi{\'c}} et~al.}{2022}]{2022LRR....25....3B}
{Bogdanovi{\'c}} T.,  {Miller} M.~C.,   {Blecha} L.,  2022, \mn@doi [Living
  Reviews in Relativity] {10.1007/s41114-022-00037-8}, \href
  {https://ui.adsabs.harvard.edu/abs/2022LRR....25....3B} {25, 3}

\bibitem[\protect\citeauthoryear{{Bonetti}, {Sesana}, {Haardt}, {Barausse}  \&
  {Colpi}}{{Bonetti} et~al.}{2019}]{Bonetti19}
{Bonetti} M.,  {Sesana} A.,  {Haardt} F.,  {Barausse} E.,   {Colpi} M.,  2019,
  \mn@doi [\mnras] {10.1093/mnras/stz903}, \href
  {https://ui.adsabs.harvard.edu/abs/2019MNRAS.486.4044B} {486, 4044}

\bibitem[\protect\citeauthoryear{{Bowen}, {Campanelli}, {Krolik}, {Mewes}  \&
  {Noble}}{{Bowen} et~al.}{2017}]{Bowen17}
{Bowen} D.~B.,  {Campanelli} M.,  {Krolik} J.~H.,  {Mewes} V.,   {Noble} S.~C.,
   2017, \mn@doi [\apj] {10.3847/1538-4357/aa63f3}, \href
  {http://adsabs.harvard.edu/abs/2017ApJ...838...42B} {838, 42}

\bibitem[\protect\citeauthoryear{{Bowen}, {Mewes}, {Campanelli}, {Noble},
  {Krolik}  \& {Zilh{\~a}o}}{{Bowen} et~al.}{2018}]{Bowen18}
{Bowen} D.~B.,  {Mewes} V.,  {Campanelli} M.,  {Noble} S.~C.,  {Krolik} J.~H.,
   {Zilh{\~a}o} M.,  2018, \mn@doi [\apjl] {10.3847/2041-8213/aaa756}, \href
  {http://adsabs.harvard.edu/abs/2018ApJ...853L..17B} {853, L17}

\bibitem[\protect\citeauthoryear{{Bowen}, {Mewes}, {Noble}, {Avara},
  {Campanelli}  \& {Krolik}}{{Bowen} et~al.}{2019}]{Bowen2019}
{Bowen} D.~B.,  {Mewes} V.,  {Noble} S.~C.,  {Avara} M.,  {Campanelli} M.,
  {Krolik} J.~H.,  2019, \mn@doi [\apj] {10.3847/1538-4357/ab2453}, \href
  {https://ui.adsabs.harvard.edu/abs/2019ApJ...879...76B} {879, 76}

\bibitem[\protect\citeauthoryear{{Breivik}, {Kremer}, {Bueno}, {Larson},
  {Coughlin}  \& {Kalogera}}{{Breivik} et~al.}{2018}]{Breivik2018}
{Breivik} K.,  {Kremer} K.,  {Bueno} M.,  {Larson} S.~L.,  {Coughlin} S.,
  {Kalogera} V.,  2018, \mn@doi [\apjl] {10.3847/2041-8213/aaaa23}, \href
  {https://ui.adsabs.harvard.edu/abs/2018ApJ...854L...1B} {854, L1}

\bibitem[\protect\citeauthoryear{{Breivik}, {Mingarelli}  \&
  {Larson}}{{Breivik} et~al.}{2020}]{Breivik2020}
{Breivik} K.,  {Mingarelli} C. M.~F.,   {Larson} S.~L.,  2020, \mn@doi [\apj]
  {10.3847/1538-4357/abab99}, \href
  {https://ui.adsabs.harvard.edu/abs/2020ApJ...901....4B} {901, 4}

\bibitem[\protect\citeauthoryear{{Capozzi} et~al.,}{{Capozzi}
  et~al.}{2017}]{capozzi17}
{Capozzi} D.,  et~al., 2017, arXiv e-prints, \href
  {https://ui.adsabs.harvard.edu/abs/2017arXiv170709066C} {p. arXiv:1707.09066}

\bibitem[\protect\citeauthoryear{{Cattorini}, {Giacomazzo}, {Haardt}  \&
  {Colpi}}{{Cattorini} et~al.}{2021}]{Cattorini21}
{Cattorini} F.,  {Giacomazzo} B.,  {Haardt} F.,   {Colpi} M.,  2021, arXiv
  e-prints, \href {https://ui.adsabs.harvard.edu/abs/2021arXiv210213166C} {p.
  arXiv:2102.13166}

\bibitem[\protect\citeauthoryear{{Cattorini}, {Maggioni}, {Giacomazzo},
  {Haardt}, {Colpi}  \& {Covino}}{{Cattorini} et~al.}{2022}]{Cattorini2022}
{Cattorini} F.,  {Maggioni} S.,  {Giacomazzo} B.,  {Haardt} F.,  {Colpi} M.,
  {Covino} S.,  2022, \mn@doi [\apjl] {10.3847/2041-8213/ac6755}, \href
  {https://ui.adsabs.harvard.edu/abs/2022ApJ...930L...1C} {930, L1}

\bibitem[\protect\citeauthoryear{{Chang}, {Strubbe}, {Menou}  \&
  {Quataert}}{{Chang} et~al.}{2010}]{Chang+10}
{Chang} P.,  {Strubbe} L.~E.,  {Menou} K.,   {Quataert} E.,  2010, \mn@doi
  [\mnras] {10.1111/j.1365-2966.2010.17056.x}, \href
  {https://ui.adsabs.harvard.edu/abs/2010MNRAS.407.2007C} {407, 2007}

\bibitem[\protect\citeauthoryear{{Colpi}}{{Colpi}}{2014}]{Colpi14}
{Colpi} M.,  2014, \mn@doi [\ssr] {10.1007/s11214-014-0067-1}, \href
  {http://adsabs.harvard.edu/abs/2014SSRv..183..189C} {183, 189}

\bibitem[\protect\citeauthoryear{{Colpi} et~al.,}{{Colpi}
  et~al.}{2019}]{Colpi19}
{Colpi} M.,  et~al., 2019, arXiv e-prints, \href
  {https://ui.adsabs.harvard.edu/abs/2019arXiv190306867C} {p. arXiv:1903.06867}

\bibitem[\protect\citeauthoryear{{Combi}, {Lopez Armengol}, {Campanelli},
  {Noble}, {Avara}, {Krolik}  \& {Bowen}}{{Combi} et~al.}{2022}]{Combi2022}
{Combi} L.,  {Lopez Armengol} F.~G.,  {Campanelli} M.,  {Noble} S.~C.,  {Avara}
  M.,  {Krolik} J.~H.,   {Bowen} D.,  2022, \mn@doi [\apj]
  {10.3847/1538-4357/ac532a10.48550/arXiv.2109.01307}, \href
  {https://ui.adsabs.harvard.edu/abs/2022ApJ...928..187C} {928, 187}

\bibitem[\protect\citeauthoryear{{Corrales}, {Haiman}  \&
  {MacFadyen}}{{Corrales} et~al.}{2010}]{Corrales+10}
{Corrales} L.~R.,  {Haiman} Z.,   {MacFadyen} A.,  2010, \mn@doi [\mnras]
  {10.1111/j.1365-2966.2010.16324.x}, \href
  {https://ui.adsabs.harvard.edu/abs/2010MNRAS.404..947C} {404, 947}

\bibitem[\protect\citeauthoryear{{Cuadra}, {Armitage}, {Alexander}  \&
  {Begelman}}{{Cuadra} et~al.}{2009}]{Cuadra2009}
{Cuadra} J.,  {Armitage} P.~J.,  {Alexander} R.~D.,   {Begelman} M.~C.,  2009,
  \mn@doi [\mnras] {10.1111/j.1365-2966.2008.14147.x}, \href
  {https://ui.adsabs.harvard.edu/abs/2009MNRAS.393.1423C} {393, 1423}

\bibitem[\protect\citeauthoryear{{D'Orazio} \& {Di Stefano}}{{D'Orazio} \& {Di
  Stefano}}{2018}]{2018MNRAS.474.2975D}
{D'Orazio} D.~J.,  {Di Stefano} R.,  2018, \mn@doi [\mnras]
  {10.1093/mnras/stx2936}, \href
  {http://adsabs.harvard.edu/abs/2018MNRAS.474.2975D} {474, 2975}

\bibitem[\protect\citeauthoryear{{Dal Canton}, Mangiagli, Noble, Schnittman,
  Ptak, Klein, Sesana  \& Camp}{{Dal Canton} et~al.}{2019}]{Dal_Canton_2019}
{Dal Canton} T.,  Mangiagli A.,  Noble S.~C.,  Schnittman J.,  Ptak A.,  Klein
  A.,  Sesana A.,   Camp J.,  2019, \mn@doi [The Astrophysical Journal]
  {10.3847/1538-4357/ab505a}, 886, 146

\bibitem[\protect\citeauthoryear{{Dayal}, {Rossi}, {Shiralilou}, {Piana},
  {Choudhury}  \& {Volonteri}}{{Dayal} et~al.}{2019}]{2019MNRAS.486.2336D}
{Dayal} P.,  {Rossi} E.~M.,  {Shiralilou} B.,  {Piana} O.,  {Choudhury} T.~R.,
   {Volonteri} M.,  2019, \mn@doi [\mnras] {10.1093/mnras/stz897}, \href
  {https://ui.adsabs.harvard.edu/abs/2019MNRAS.486.2336D} {486, 2336}

\bibitem[\protect\citeauthoryear{{De Rosa} et~al.,}{{De Rosa}
  et~al.}{2019}]{DeRosa2019}
{De Rosa} A.,  et~al., 2019, \mn@doi [\nar] {10.1016/j.newar.2020.101525},
  \href {https://ui.adsabs.harvard.edu/abs/2019NewAR..8601525D} {86, 101525}

\bibitem[\protect\citeauthoryear{{Deason}, {Wetzel}  \&
  {Garrison-Kimmel}}{{Deason} et~al.}{2014}]{Deason14}
{Deason} A.,  {Wetzel} A.,   {Garrison-Kimmel} S.,  2014, \mn@doi [\apj]
  {10.1088/0004-637X/794/2/115}, \href
  {https://ui.adsabs.harvard.edu/abs/2014ApJ...794..115D} {794, 115}

\bibitem[\protect\citeauthoryear{{Derdzinski}, {D'Orazio}, {Duffell}, {Haiman}
  \& {MacFadyen}}{{Derdzinski} et~al.}{2021}]{Derdzinski21}
{Derdzinski} A.,  {D'Orazio} D.,  {Duffell} P.,  {Haiman} Z.,   {MacFadyen} A.,
   2021, \mn@doi [\mnras] {10.1093/mnras/staa3976}, \href
  {https://ui.adsabs.harvard.edu/abs/2021MNRAS.501.3540D} {501, 3540}

\bibitem[\protect\citeauthoryear{{Duffell}, {D'Orazio}, {Derdzinski}, {Haiman},
  {MacFadyen}, {Rosen}  \& {Zrake}}{{Duffell} et~al.}{2020}]{Duffell2020}
{Duffell} P.~C.,  {D'Orazio} D.,  {Derdzinski} A.,  {Haiman} Z.,  {MacFadyen}
  A.,  {Rosen} A.~L.,   {Zrake} J.,  2020, \mn@doi [\apj]
  {10.3847/1538-4357/abab95}, \href
  {https://ui.adsabs.harvard.edu/abs/2020ApJ...901...25D} {901, 25}

\bibitem[\protect\citeauthoryear{{Fabian} et~al.,}{{Fabian}
  et~al.}{2009}]{fabian09b}
{Fabian} A.~C.,  et~al., 2009, in astro2010: The Astronomy and Astrophysics
  Decadal Survey.  (\mn@eprint {arXiv} {0903.4424})

\bibitem[\protect\citeauthoryear{{Farris}, {Duffell}, {MacFadyen}  \&
  {Haiman}}{{Farris} et~al.}{2014}]{Farris14}
{Farris} B.~D.,  {Duffell} P.,  {MacFadyen} A.~I.,   {Haiman} Z.,  2014,
  \mn@doi [\apj] {10.1088/0004-637X/783/2/134}, \href
  {http://adsabs.harvard.edu/abs/2014ApJ...783..134F} {783, 134}

\bibitem[\protect\citeauthoryear{{Farris}, {Duffell}, {MacFadyen}  \&
  {Haiman}}{{Farris} et~al.}{2015}]{Farris2015}
{Farris} B.~D.,  {Duffell} P.,  {MacFadyen} A.~I.,   {Haiman} Z.,  2015,
  \mn@doi [\mnras] {10.1093/mnrasl/slu184}, \href
  {https://ui.adsabs.harvard.edu/abs/2015MNRAS.447L..80F} {447, L80}

\bibitem[\protect\citeauthoryear{{Gallo} \& {Sesana}}{{Gallo} \&
  {Sesana}}{2019}]{Gallo2019}
{Gallo} E.,  {Sesana} A.,  2019, \mn@doi [\apjl] {10.3847/2041-8213/ab40c6},
  \href {https://ui.adsabs.harvard.edu/abs/2019ApJ...883L..18G} {883, L18}

\bibitem[\protect\citeauthoryear{{Georgakakis}, {Nandra}, {Laird}, {Aird}  \&
  {Trichas}}{{Georgakakis} et~al.}{2008}]{Georgakakis08}
{Georgakakis} A.,  {Nandra} K.,  {Laird} E.~S.,  {Aird} J.,   {Trichas} M.,
  2008, \mn@doi [\mnras] {10.1111/j.1365-2966.2008.13423.x}, \href
  {https://ui.adsabs.harvard.edu/abs/2008MNRAS.388.1205G} {388, 1205}

\bibitem[\protect\citeauthoryear{{Ghirlanda} et~al.,}{{Ghirlanda}
  et~al.}{2019}]{2019Sci...363..968G}
{Ghirlanda} G.,  et~al., 2019, \mn@doi [Science] {10.1126/science.aau8815},
  \href {https://ui.adsabs.harvard.edu/abs/2019Sci...363..968G} {363, 968}

\bibitem[\protect\citeauthoryear{{Giacomazzo}, {Baker}, {Miller}, {Reynolds}
  \& {van Meter}}{{Giacomazzo} et~al.}{2012}]{Giacomazzo12}
{Giacomazzo} B.,  {Baker} J.~G.,  {Miller} M.~C.,  {Reynolds} C.~S.,   {van
  Meter} J.~R.,  2012, \mn@doi [\apjl] {10.1088/2041-8205/752/1/L15}, \href
  {https://ui.adsabs.harvard.edu/abs/2012ApJ...752L..15G} {752, L15}

\bibitem[\protect\citeauthoryear{{Gold}, {Paschalidis}, {Ruiz}, {Shapiro},
  {Etienne}  \& {Pfeiffer}}{{Gold} et~al.}{2014}]{Gold2014}
{Gold} R.,  {Paschalidis} V.,  {Ruiz} M.,  {Shapiro} S.~L.,  {Etienne} Z.~B.,
  {Pfeiffer} H.~P.,  2014, \mn@doi [\prd] {10.1103/PhysRevD.90.104030}, \href
  {https://ui.adsabs.harvard.edu/abs/2014PhRvD..90j4030G} {90, 104030}

\bibitem[\protect\citeauthoryear{{Greene} \& {Ho}}{{Greene} \&
  {Ho}}{2007}]{Greene2007}
{Greene} J.~E.,  {Ho} L.~C.,  2007, \mn@doi [\apj] {10.1086/522082}, \href
  {https://ui.adsabs.harvard.edu/abs/2007ApJ...670...92G} {670, 92}

\bibitem[\protect\citeauthoryear{{Greene}, {Strader}  \& {Ho}}{{Greene}
  et~al.}{2020}]{Greene20}
{Greene} J.~E.,  {Strader} J.,   {Ho} L.~C.,  2020, \mn@doi [\araa]
  {10.1146/annurev-astro-032620-021835}, \href
  {https://ui.adsabs.harvard.edu/abs/2020ARA&A..58..257G} {58, 257}

\bibitem[\protect\citeauthoryear{{Guti{\'e}rrez}, {Combi}, {Noble},
  {Campanelli}, {Krolik}, {L{\'o}pez Armengol}  \&
  {Garc{\'\i}a}}{{Guti{\'e}rrez} et~al.}{2022}]{Gutierrez2022}
{Guti{\'e}rrez} E.~M.,  {Combi} L.,  {Noble} S.~C.,  {Campanelli} M.,  {Krolik}
  J.~H.,  {L{\'o}pez Armengol} F.,   {Garc{\'\i}a} F.,  2022, \mn@doi [\apj]
  {10.3847/1538-4357/ac56de10.48550/arXiv.2112.09773}, \href
  {https://ui.adsabs.harvard.edu/abs/2022ApJ...928..137G} {928, 137}

\bibitem[\protect\citeauthoryear{{Haiman}}{{Haiman}}{2017}]{Haiman17}
{Haiman} Z.,  2017, \mn@doi [\prd] {10.1103/PhysRevD.96.023004}, \href
  {http://adsabs.harvard.edu/abs/2017PhRvD..96b3004H} {96, 023004}

\bibitem[\protect\citeauthoryear{{Haiman}, {Kocsis}  \& {Menou}}{{Haiman}
  et~al.}{2009}]{Haiman09}
{Haiman} Z.,  {Kocsis} B.,   {Menou} K.,  2009, \mn@doi [\apj]
  {10.1088/0004-637X/700/2/1952}, \href
  {http://adsabs.harvard.edu/abs/2009ApJ...700.1952H} {700, 1952}

\bibitem[\protect\citeauthoryear{Hallinan et~al.,}{Hallinan
  et~al.}{2017}]{Hallinan17}
Hallinan G.,  et~al., 2017, \mn@doi [Science] {10.1126/science.aap9855}, 358,
  1579

\bibitem[\protect\citeauthoryear{{Hopkins}, {Hernquist}, {Cox}, {Di Matteo},
  {Martini}, {Robertson}  \& {Springel}}{{Hopkins} et~al.}{2005}]{hopkins05}
{Hopkins} P.~F.,  {Hernquist} L.,  {Cox} T.~J.,  {Di Matteo} T.,  {Martini} P.,
   {Robertson} B.,   {Springel} V.,  2005, \mn@doi [\apj] {10.1086/432438},
  \href {http://adsabs.harvard.edu/abs/2005ApJ...630..705H} {630, 705}

\bibitem[\protect\citeauthoryear{{Ilbert} et~al.,}{{Ilbert}
  et~al.}{2013}]{Ilbert13}
{Ilbert} O.,  et~al., 2013, \mn@doi [\aap] {10.1051/0004-6361/201321100}, \href
  {https://ui.adsabs.harvard.edu/\#abs/2013A&A...556A..55I} {556, A55}

\bibitem[\protect\citeauthoryear{{Katz}, {Marsat}, {Chua}, {Babak}  \&
  {Larson}}{{Katz} et~al.}{2020}]{Katz20}
{Katz} M.~L.,  {Marsat} S.,  {Chua} A. J.~K.,  {Babak} S.,   {Larson} S.~L.,
  2020, \mn@doi [Phys. Rev. D] {10.1103/PhysRevD.102.023033}, \href
  {https://ui.adsabs.harvard.edu/abs/2020PhRvD.102b3033K} {102, 023033}

\bibitem[\protect\citeauthoryear{{Kelly}, {Baker}, {Etienne}, {Giacomazzo}  \&
  {Schnittman}}{{Kelly} et~al.}{2017}]{Kelly17}
{Kelly} B.~J.,  {Baker} J.~G.,  {Etienne} Z.~B.,  {Giacomazzo} B.,
  {Schnittman} J.,  2017, \mn@doi [\prd] {10.1103/PhysRevD.96.123003}, \href
  {http://adsabs.harvard.edu/abs/2017PhRvD..96l3003K} {96, 123003}

\bibitem[\protect\citeauthoryear{{Khan}, {Paschalidis}, {Ruiz}  \&
  {Shapiro}}{{Khan} et~al.}{2018}]{Khan18}
{Khan} A.,  {Paschalidis} V.,  {Ruiz} M.,   {Shapiro} S.~L.,  2018, \mn@doi
  [\prd] {10.1103/PhysRevD.97.044036}, \href
  {http://adsabs.harvard.edu/abs/2018PhRvD..97d4036K} {97, 044036}

\bibitem[\protect\citeauthoryear{{Kocsis}, {Yunes}  \& {Loeb}}{{Kocsis}
  et~al.}{2011}]{2011PhRvD..84b4032K}
{Kocsis} B.,  {Yunes} N.,   {Loeb} A.,  2011, \mn@doi [\prd]
  {10.1103/PhysRevD.84.024032}, \href
  {https://ui.adsabs.harvard.edu/abs/2011PhRvD..84b4032K} {84, 024032}

\bibitem[\protect\citeauthoryear{{Kormendy} \& {Ho}}{{Kormendy} \&
  {Ho}}{2013}]{KormendyHo2013}
{Kormendy} J.,  {Ho} L.~C.,  2013, \mn@doi [\araa]
  {10.1146/annurev-astro-082708-101811}, \href
  {http://cdsads.u-strasbg.fr/abs/2013ARA} {51, 511}

\bibitem[\protect\citeauthoryear{{Korol} et~al.,}{{Korol}
  et~al.}{2020}]{Korol2020}
{Korol} V.,  et~al., 2020, \mn@doi [\aap] {10.1051/0004-6361/202037764}, \href
  {https://ui.adsabs.harvard.edu/abs/2020A&A...638A.153K} {638, A153}

\bibitem[\protect\citeauthoryear{{Koss} et~al.,}{{Koss} et~al.}{2018}]{koss18}
{Koss} M.~J.,  et~al., 2018, \mn@doi [\nat] {10.1038/s41586-018-0652-7}, \href
  {http://adsabs.harvard.edu/abs/2018Natur.563..214K} {563, 214}

\bibitem[\protect\citeauthoryear{{Kremer}, {Breivik}, {Larson}  \&
  {Kalogera}}{{Kremer} et~al.}{2017}]{Kremer2017}
{Kremer} K.,  {Breivik} K.,  {Larson} S.~L.,   {Kalogera} V.,  2017, \mn@doi
  [\apj] {10.3847/1538-4357/aa8557}, \href
  {https://ui.adsabs.harvard.edu/abs/2017ApJ...846...95K} {846, 95}

\bibitem[\protect\citeauthoryear{LISA{\ }Science{\ }Study{\ }Team}{LISA{\
  }Science{\ }Study{\ }Team}{2018}]{lisa_requirement_docs}
LISA{\ }Science{\ }Study{\ }Team 2018, Technical Report ESA-L3-EST-SCI-RS-001.
  European Space Agency

\bibitem[\protect\citeauthoryear{{Lang} \& {Hughes}}{{Lang} \&
  {Hughes}}{2006}]{2006PhRvD..74l2001L}
{Lang} R.~N.,  {Hughes} S.~A.,  2006, \mn@doi [\prd]
  {10.1103/PhysRevD.74.122001}, \href
  {https://ui.adsabs.harvard.edu/abs/2006PhRvD..74l2001L} {74, 122001}

\bibitem[\protect\citeauthoryear{{Lippai}, {Frei}  \& {Haiman}}{{Lippai}
  et~al.}{2008}]{Lippai+08}
{Lippai} Z.,  {Frei} Z.,   {Haiman} Z.,  2008, \mn@doi [\apjl]
  {10.1086/587034}, \href
  {https://ui.adsabs.harvard.edu/abs/2008ApJ...676L...5L} {676, L5}

\bibitem[\protect\citeauthoryear{{Lodato}, {Nayakshin}, {King}  \&
  {Pringle}}{{Lodato} et~al.}{2009}]{Lodato2009}
{Lodato} G.,  {Nayakshin} S.,  {King} A.~R.,   {Pringle} J.~E.,  2009, \mn@doi
  [\mnras] {10.1111/j.1365-2966.2009.15179.x}, \href
  {https://ui.adsabs.harvard.edu/abs/2009MNRAS.398.1392L} {398, 1392}

\bibitem[\protect\citeauthoryear{{London} et~al.,}{{London}
  et~al.}{2018}]{London18}
{London} L.,  et~al., 2018, \mn@doi [Phys. Rev. Letters]
  {10.1103/PhysRevLett.120.161102}, \href
  {https://ui.adsabs.harvard.edu/abs/2018PhRvL.120p1102L} {120, 161102}

\bibitem[\protect\citeauthoryear{{Lops}, {Izquierdo-Villalba}, {Colpi},
  {Bonoli}, {Sesana}  \& {Mangiagli}}{{Lops} et~al.}{2022}]{Lops2022}
{Lops} G.,  {Izquierdo-Villalba} D.,  {Colpi} M.,  {Bonoli} S.,  {Sesana} A.,
  {Mangiagli} A.,  2022, arXiv e-prints, \href
  {https://ui.adsabs.harvard.edu/abs/2022arXiv220710683L} {p. arXiv:2207.10683}

\bibitem[\protect\citeauthoryear{{Lousto}, {Campanelli}, {Zlochower}  \&
  {Nakano}}{{Lousto} et~al.}{2010}]{Lousto10}
{Lousto} C.~O.,  {Campanelli} M.,  {Zlochower} Y.,   {Nakano} H.,  2010,
  \mn@doi [Classical and Quantum Gravity] {10.1088/0264-9381/27/11/114006},
  \href {https://ui.adsabs.harvard.edu/abs/2010CQGra..27k4006L} {27, 114006}

\bibitem[\protect\citeauthoryear{{Lusso} et~al.,}{{Lusso}
  et~al.}{2012}]{lusso12}
{Lusso} E.,  et~al., 2012, \mn@doi [\mnras] {10.1111/j.1365-2966.2012.21513.x},
  \href {http://adsabs.harvard.edu/abs/2012MNRAS.425..623L} {425, 623}

\bibitem[\protect\citeauthoryear{{Man}, {Zirm}  \& {Toft}}{{Man}
  et~al.}{2016}]{Man16}
{Man} A. W.~S.,  {Zirm} A.~W.,   {Toft} S.,  2016, \mn@doi [\apj]
  {10.3847/0004-637X/830/2/89}, \href
  {https://ui.adsabs.harvard.edu/abs/2016ApJ...830...89M} {830, 89}

\bibitem[\protect\citeauthoryear{{Mangiagli} et~al.,}{{Mangiagli}
  et~al.}{2020}]{Mangiagli20}
{Mangiagli} A.,  et~al., 2020, \mn@doi [\prd] {10.1103/PhysRevD.102.084056},
  \href {https://ui.adsabs.harvard.edu/abs/2020PhRvD.102h4056M} {102, 084056}

\bibitem[\protect\citeauthoryear{{Mangiagli}, {Caprini}, {Volonteri}, {Marsat},
  {Vergani}, {Tamanini}  \& {Inchausp{\'e}}}{{Mangiagli}
  et~al.}{2022}]{Mangiagli2022}
{Mangiagli} A.,  {Caprini} C.,  {Volonteri} M.,  {Marsat} S.,  {Vergani} S.,
  {Tamanini} N.,   {Inchausp{\'e}} H.,  2022, arXiv e-prints, \href
  {https://ui.adsabs.harvard.edu/abs/2022arXiv220710678M} {p. arXiv:2207.10678}

\bibitem[\protect\citeauthoryear{{Maoz}, {Hallakoun}  \& {Badenes}}{{Maoz}
  et~al.}{2018}]{maoz18}
{Maoz} D.,  {Hallakoun} N.,   {Badenes} C.,  2018, \mn@doi [\mnras]
  {10.1093/mnras/sty339}, \href
  {http://adsabs.harvard.edu/abs/2018MNRAS.476.2584M} {476, 2584}

\bibitem[\protect\citeauthoryear{{Marsat}, {Baker}  \& {Dal Canton}}{{Marsat}
  et~al.}{2020}]{Marsat20}
{Marsat} S.,  {Baker} J.~G.,   {Dal Canton} T.,  2020, arXiv e-prints, \href
  {https://ui.adsabs.harvard.edu/abs/2020arXiv200300357M} {p. arXiv:2003.00357}

\bibitem[\protect\citeauthoryear{{McKernan}, {Ford}, {Kocsis}, {Lyra}  \&
  {Winter}}{{McKernan} et~al.}{2014}]{2014MNRAS.441..900M}
{McKernan} B.,  {Ford} K.~E.~S.,  {Kocsis} B.,  {Lyra} W.,   {Winter} L.~M.,
  2014, \mn@doi [\mnras] {10.1093/mnras/stu553}, \href
  {https://ui.adsabs.harvard.edu/abs/2014MNRAS.441..900M} {441, 900}

\bibitem[\protect\citeauthoryear{{Megevand}, {Anderson}, {Frank}, {Hirschmann},
  {Lehner}, {Liebling}, {Motl}  \& {Neilsen}}{{Megevand}
  et~al.}{2009}]{megevand+09}
{Megevand} M.,  {Anderson} M.,  {Frank} J.,  {Hirschmann} E.~W.,  {Lehner} L.,
  {Liebling} S.~L.,  {Motl} P.~M.,   {Neilsen} D.,  2009, \mn@doi [\prd]
  {10.1103/PhysRevD.80.024012}, \href
  {https://ui.adsabs.harvard.edu/abs/2009PhRvD..80b4012M} {80, 024012}

\bibitem[\protect\citeauthoryear{{Meidinger}, {Albrecht}, {Bonholzer},
  {M{\"u}ller-Seidlitz}, {Nandra}, {Ott}, {Plattner}  \&
  {Treberspurg}}{{Meidinger} et~al.}{2019}]{meidinger19}
{Meidinger} N.,  {Albrecht} S.,  {Bonholzer} M.,  {M{\"u}ller-Seidlitz} J.,
  {Nandra} K.,  {Ott} S.,  {Plattner} M.,   {Treberspurg} W.,  2019, in UV,
  X-Ray, and Gamma-Ray Space Instrumentation for Astronomy XXI. p. 111180Y,
  \mn@doi{10.1117/12.2528109}

\bibitem[\protect\citeauthoryear{{Merloni} et~al.,}{{Merloni}
  et~al.}{2012}]{merloni12}
{Merloni} A.,  et~al., 2012, arXiv e-prints, \href
  {http://adsabs.harvard.edu/abs/2012arXiv1209.3114M} {}

\bibitem[\protect\citeauthoryear{{Metzger}}{{Metzger}}{2019}]{Metzger20}
{Metzger} B.~D.,  2019, \mn@doi [Living Reviews in Relativity]
  {10.1007/s41114-019-0024-0}, \href
  {https://ui.adsabs.harvard.edu/abs/2019LRR....23....1M} {23, 1}

\bibitem[\protect\citeauthoryear{{Milosavljevi{\'c}} \&
  {Phinney}}{{Milosavljevi{\'c}} \& {Phinney}}{2005}]{Milosavljevic2005}
{Milosavljevi{\'c}} M.,  {Phinney} E.~S.,  2005, \mn@doi [\apjl]
  {10.1086/429618}, \href
  {https://ui.adsabs.harvard.edu/abs/2005ApJ...622L..93M} {622, L93}

\bibitem[\protect\citeauthoryear{{Miniutti}, {Ponti}, {Greene}, {Ho}, {Fabian}
  \& {Iwasawa}}{{Miniutti} et~al.}{2009}]{miniutti09}
{Miniutti} G.,  {Ponti} G.,  {Greene} J.~E.,  {Ho} L.~C.,  {Fabian} A.~C.,
  {Iwasawa} K.,  2009, \mn@doi [\mnras] {10.1111/j.1365-2966.2008.14334.x},
  \href {http://adsabs.harvard.edu/abs/2009MNRAS.394..443M} {394, 443}

\bibitem[\protect\citeauthoryear{{Moody}, {Shi}  \& {Stone}}{{Moody}
  et~al.}{2019}]{Moody2019}
{Moody} M. S.~L.,  {Shi} J.-M.,   {Stone} J.~M.,  2019, \mn@doi [\apj]
  {10.3847/1538-4357/ab09ee}, \href
  {https://ui.adsabs.harvard.edu/abs/2019ApJ...875...66M} {875, 66}

\bibitem[\protect\citeauthoryear{{Mu{\~n}oz}, {Miranda}  \& {Lai}}{{Mu{\~n}oz}
  et~al.}{2019}]{Munoz-Lai2019}
{Mu{\~n}oz} D.~J.,  {Miranda} R.,   {Lai} D.,  2019, \mn@doi [\apj]
  {10.3847/1538-4357/aaf867}, \href
  {https://ui.adsabs.harvard.edu/abs/2019ApJ...871...84M} {871, 84}

\bibitem[\protect\citeauthoryear{{Nandra} et~al.,}{{Nandra}
  et~al.}{2013}]{nandra13}
{Nandra} K.,  et~al., 2013, \mn@doi [arXiv e-prints]
  {10.48550/arXiv.1306.2307}, \href
  {https://ui.adsabs.harvard.edu/abs/2013arXiv1306.2307N} {p. arXiv:1306.2307}

\bibitem[\protect\citeauthoryear{{Nelemans}, {Yungelson}  \& {Portegies
  Zwart}}{{Nelemans} et~al.}{2001}]{Nelemans2001}
{Nelemans} G.,  {Yungelson} L.~R.,   {Portegies Zwart} S.~F.,  2001, \mn@doi
  [\aap] {10.1051/0004-6361:20010683}, \href
  {https://ui.adsabs.harvard.edu/abs/2001A&A...375..890N} {375, 890}

\bibitem[\protect\citeauthoryear{{O'Leary}, {Moster}, {Naab}  \&
  {Somerville}}{{O'Leary} et~al.}{2021}]{Oleary21}
{O'Leary} J.~A.,  {Moster} B.~P.,  {Naab} T.,   {Somerville} R.~S.,  2021,
  \mn@doi [\mnras] {10.1093/mnras/staa3746}, \href
  {https://ui.adsabs.harvard.edu/abs/2021MNRAS.501.3215O} {501, 3215}

\bibitem[\protect\citeauthoryear{{O'Neill}, {Miller}, {Bogdanovi{\'c}},
  {Reynolds}  \& {Schnittman}}{{O'Neill} et~al.}{2009}]{Neill+09}
{O'Neill} S.~M.,  {Miller} M.~C.,  {Bogdanovi{\'c}} T.,  {Reynolds} C.~S.,
  {Schnittman} J.~D.,  2009, \mn@doi [\apj] {10.1088/0004-637X/700/1/859},
  \href {https://ui.adsabs.harvard.edu/abs/2009ApJ...700..859O} {700, 859}

\bibitem[\protect\citeauthoryear{{Panessa}, {Barcons}, {Bassani}, {Cappi},
  {Carrera}, {Ho}  \& {Pellegrini}}{{Panessa} et~al.}{2007}]{panessa07}
{Panessa} F.,  {Barcons} X.,  {Bassani} L.,  {Cappi} M.,  {Carrera} F.~J.,
  {Ho} L.~C.,   {Pellegrini} S.,  2007, \mn@doi [\aap]
  {10.1051/0004-6361:20066943}, \href
  {http://adsabs.harvard.edu/abs/2007A%26A...467..519P} {467, 519}

\bibitem[\protect\citeauthoryear{{Paschalidis}, {Bright}, {Ruiz}  \&
  {Gold}}{{Paschalidis} et~al.}{2021}]{2021ApJ...910L..26P}
{Paschalidis} V.,  {Bright} J.,  {Ruiz} M.,   {Gold} R.,  2021, \mn@doi [\apjl]
  {10.3847/2041-8213/abee21}, \href
  {https://ui.adsabs.harvard.edu/abs/2021ApJ...910L..26P} {910, L26}

\bibitem[\protect\citeauthoryear{{Peres}}{{Peres}}{1962}]{Peres62}
{Peres} A.,  1962, \mn@doi [Physical Review] {10.1103/PhysRev.128.2471}, \href
  {https://ui.adsabs.harvard.edu/abs/1962PhRv..128.2471P} {128, 2471}

\bibitem[\protect\citeauthoryear{{Piro} et~al.,}{{Piro} et~al.}{2022}]{Piro22}
{Piro} L.,  et~al., 2022, \mn@doi [Experimental Astronomy]
  {10.1007/s10686-022-09865-6}, \href
  {https://ui.adsabs.harvard.edu/abs/2022ExA....54...23P} {54, 23}

\bibitem[\protect\citeauthoryear{{Ponti}, {Papadakis}, {Bianchi}, {Guainazzi},
  {Matt}, {Uttley}  \& {Bonilla}}{{Ponti} et~al.}{2012}]{ponti12}
{Ponti} G.,  {Papadakis} I.,  {Bianchi} S.,  {Guainazzi} M.,  {Matt} G.,
  {Uttley} P.,   {Bonilla} N.~F.,  2012, \mn@doi [\aap]
  {10.1051/0004-6361/201118326}, \href
  {http://adsabs.harvard.edu/abs/2012A%26A...542A..83P} {542, A83}

\bibitem[\protect\citeauthoryear{{Pratten}, {Klein}, {Moore}, {Middleton},
  {Steinle}, {Schmidt}  \& {Vecchio}}{{Pratten} et~al.}{2022}]{Pratten2022}
{Pratten} G.,  {Klein} A.,  {Moore} C.~J.,  {Middleton} H.,  {Steinle} N.,
  {Schmidt} P.,   {Vecchio} A.,  2022, \mn@doi [arXiv e-prints]
  {10.48550/arXiv.2212.02572}, \href
  {https://ui.adsabs.harvard.edu/abs/2022arXiv221202572P} {p. arXiv:2212.02572}

\bibitem[\protect\citeauthoryear{{Predehl} et~al.,}{{Predehl}
  et~al.}{2021}]{predehl21}
{Predehl} P.,  et~al., 2021, \mn@doi [\aap] {10.1051/0004-6361/202039313},
  \href {https://ui.adsabs.harvard.edu/abs/2021A&A...647A...1P} {647, A1}

\bibitem[\protect\citeauthoryear{{Rau} et~al.,}{{Rau} et~al.}{2016}]{Rau16}
{Rau} A.,  et~al., 2016, in {den Herder} J.-W.~A.,  {Takahashi} T.,   {Bautz}
  M.,  eds,  Society of Photo-Optical Instrumentation Engineers (SPIE)
  Conference Series Vol. 9905, Space Telescopes and Instrumentation 2016:
  Ultraviolet to Gamma Ray. p. 99052B (\mn@eprint {arXiv} {1607.00878}),
  \mn@doi{10.1117/12.2235268}

\bibitem[\protect\citeauthoryear{{Reines} \& {Comastri}}{{Reines} \&
  {Comastri}}{2016}]{Reines-Comastri2016}
{Reines} A.~E.,  {Comastri} A.,  2016, \mn@doi [\pasa] {10.1017/pasa.2016.46},
  \href {https://ui.adsabs.harvard.edu/abs/2016PASA...33...54R} {33, e054}

\bibitem[\protect\citeauthoryear{{Reines}, {Condon}, {Darling}  \&
  {Greene}}{{Reines} et~al.}{2019}]{Reines2019}
{Reines} A.,  {Condon} J.,  {Darling} J.~K.,   {Greene} J.,  2019, in American
  Astronomical Society Meeting Abstracts \#233. p. 134.02

\bibitem[\protect\citeauthoryear{{Ricci} et~al.,}{{Ricci}
  et~al.}{2017}]{ricci17}
{Ricci} C.,  et~al., 2017, \mn@doi [\nat] {10.1038/nature23906}, \href
  {http://adsabs.harvard.edu/abs/2017Natur.549..488R} {549, 488}

\bibitem[\protect\citeauthoryear{{Ricci} et~al.,}{{Ricci}
  et~al.}{2020}]{Ricci20}
{Ricci} C.,  et~al., 2020, \mn@doi [\apjl] {10.3847/2041-8213/ab91a1}, \href
  {https://ui.adsabs.harvard.edu/abs/2020ApJ...898L...1R} {898, L1}

\bibitem[\protect\citeauthoryear{{Robson}, {Cornish}  \& {Liu}}{{Robson}
  et~al.}{2019}]{Robson19}
{Robson} T.,  {Cornish} N.~J.,   {Liu} C.,  2019, \mn@doi [Classical and
  Quantum Gravity] {10.1088/1361-6382/ab1101}, \href
  {https://ui.adsabs.harvard.edu/abs/2019CQGra..36j5011R} {36, 105011}

\bibitem[\protect\citeauthoryear{{Roedig}, {Sesana}, {Dotti}, {Cuadra},
  {Amaro-Seoane}  \& {Haardt}}{{Roedig} et~al.}{2012}]{Roedig12}
{Roedig} C.,  {Sesana} A.,  {Dotti} M.,  {Cuadra} J.,  {Amaro-Seoane} P.,
  {Haardt} F.,  2012, \mn@doi [\aap] {10.1051/0004-6361/201219986}, \href
  {https://ui.adsabs.harvard.edu/abs/2012A&A...545A.127R} {545, A127}

\bibitem[\protect\citeauthoryear{{Roedig}, {Krolik}  \& {Miller}}{{Roedig}
  et~al.}{2014}]{Roedig14}
{Roedig} C.,  {Krolik} J.~H.,   {Miller} M.~C.,  2014, \mn@doi [\apj]
  {10.1088/0004-637X/785/2/115}, \href
  {http://adsabs.harvard.edu/abs/2014ApJ...785..115R} {785, 115}

\bibitem[\protect\citeauthoryear{{Rosotti}, {Lodato}  \& {Price}}{{Rosotti}
  et~al.}{2012}]{RosottiLodato12}
{Rosotti} G.~P.,  {Lodato} G.,   {Price} D.~J.,  2012, \mn@doi [\mnras]
  {10.1111/j.1365-2966.2012.21488.x}, \href
  {https://ui.adsabs.harvard.edu/abs/2012MNRAS.425.1958R} {425, 1958}

\bibitem[\protect\citeauthoryear{{Rossi}, {Lodato}, {Armitage}, {Pringle}  \&
  {King}}{{Rossi} et~al.}{2010}]{Rossi10}
{Rossi} E.~M.,  {Lodato} G.,  {Armitage} P.~J.,  {Pringle} J.~E.,   {King}
  A.~R.,  2010, \mn@doi [\mnras] {10.1111/j.1365-2966.2009.15802.x}, \href
  {http://adsabs.harvard.edu/abs/2010MNRAS.401.2021R} {401, 2021}

\bibitem[\protect\citeauthoryear{{Ryan}, {van Eerten}, {Piro}  \&
  {Troja}}{{Ryan} et~al.}{2020}]{Ryan20}
{Ryan} G.,  {van Eerten} H.,  {Piro} L.,   {Troja} E.,  2020, \mn@doi [\apj]
  {10.3847/1538-4357/ab93cf}, \href
  {https://ui.adsabs.harvard.edu/abs/2020ApJ...896..166R} {896, 166}

\bibitem[\protect\citeauthoryear{{Saini}, {Bhat}  \& {Arun}}{{Saini}
  et~al.}{2022}]{Saini22}
{Saini} P.,  {Bhat} S.~A.,   {Arun} K.~G.,  2022, \mn@doi [\prd]
  {10.1103/PhysRevD.106.10401510.48550/arXiv.2208.03004}, \href
  {https://ui.adsabs.harvard.edu/abs/2022PhRvD.106j4015S} {106, 104015}

\bibitem[\protect\citeauthoryear{{Schnittman} \& {Krolik}}{{Schnittman} \&
  {Krolik}}{2008}]{Schnittman2008}
{Schnittman} J.~D.,  {Krolik} J.~H.,  2008, \mn@doi [\apj] {10.1086/590363},
  \href {https://ui.adsabs.harvard.edu/abs/2008ApJ...684..835S} {684, 835}

\bibitem[\protect\citeauthoryear{{Schutz}}{{Schutz}}{1986}]{1986Natur.323..310S}
{Schutz} B.~F.,  1986, \mn@doi [\nat] {10.1038/323310a0}, \href
  {https://ui.adsabs.harvard.edu/abs/1986Natur.323..310S} {323, 310}

\bibitem[\protect\citeauthoryear{{Serafinelli} et~al.,}{{Serafinelli}
  et~al.}{2020}]{Serafinelli2020}
{Serafinelli} R.,  et~al., 2020, \mn@doi [\apj] {10.3847/1538-4357/abb3c3},
  \href {https://ui.adsabs.harvard.edu/abs/2020ApJ...902...10S} {902, 10}

\bibitem[\protect\citeauthoryear{{Sesana}}{{Sesana}}{2016}]{Sesana2016-LIGO-LISA-BHs}
{Sesana} A.,  2016, \mn@doi [Physical Review Letters]
  {10.1103/PhysRevLett.116.231102}, \href
  {http://adsabs.harvard.edu/abs/2016PhRvL.116w1102S} {116, 231102}

\bibitem[\protect\citeauthoryear{{Sesana}, {Gair}, {Berti}  \&
  {Volonteri}}{{Sesana} et~al.}{2011}]{2011PhRvD..83d4036S}
{Sesana} A.,  {Gair} J.,  {Berti} E.,   {Volonteri} M.,  2011, \mn@doi [\prd]
  {10.1103/PhysRevD.83.044036}, \href
  {http://adsabs.harvard.edu/abs/2011PhRvD..83d4036S} {83, 044036}

\bibitem[\protect\citeauthoryear{{Severgnini} et~al.,}{{Severgnini}
  et~al.}{2018}]{Severgnini2018}
{Severgnini} P.,  et~al., 2018, \mn@doi [\mnras] {10.1093/mnras/sty1699}, \href
  {https://ui.adsabs.harvard.edu/abs/2018MNRAS.479.3804S} {479, 3804}

\bibitem[\protect\citeauthoryear{{Shields} \& {Bonning}}{{Shields} \&
  {Bonning}}{2008}]{Shields+Bonning08}
{Shields} G.~A.,  {Bonning} E.~W.,  2008, \mn@doi [\apj] {10.1086/589427},
  \href {https://ui.adsabs.harvard.edu/abs/2008ApJ...682..758S} {682, 758}

\bibitem[\protect\citeauthoryear{{Smail}, {Hogg}, {Yan}  \& {Cohen}}{{Smail}
  et~al.}{1995}]{Smail95}
{Smail} I.,  {Hogg} D.~W.,  {Yan} L.,   {Cohen} J.~G.,  1995, \mn@doi [\apjl]
  {10.1086/309647}, \href
  {https://ui.adsabs.harvard.edu/abs/1995ApJ...449L.105S} {449, L105}

\bibitem[\protect\citeauthoryear{{Sukov{\'a}}, {Zaja{\v{c}}ek}, {Witzany}  \&
  {Karas}}{{Sukov{\'a}} et~al.}{2021}]{2021ApJ...917...43S}
{Sukov{\'a}} P.,  {Zaja{\v{c}}ek} M.,  {Witzany} V.,   {Karas} V.,  2021,
  \mn@doi [\apj] {10.3847/1538-4357/ac05c610.48550/arXiv.2102.08135}, \href
  {https://ui.adsabs.harvard.edu/abs/2021ApJ...917...43S} {917, 43}

\bibitem[\protect\citeauthoryear{{Tamanini}, {Caprini}, {Barausse}, {Sesana},
  {Klein}  \& {Petiteau}}{{Tamanini} et~al.}{2016}]{Tamanini16}
{Tamanini} N.,  {Caprini} C.,  {Barausse} E.,  {Sesana} A.,  {Klein} A.,
  {Petiteau} A.,  2016, \mn@doi [\jcap] {10.1088/1475-7516/2016/04/002}, \href
  {http://adsabs.harvard.edu/abs/2016JCAP...04..002T} {4, 002}

\bibitem[\protect\citeauthoryear{{Tanaka} \& {Menou}}{{Tanaka} \&
  {Menou}}{2010}]{Tanaka2010}
{Tanaka} T.,  {Menou} K.,  2010, \mn@doi [\apj] {10.1088/0004-637X/714/1/404},
  \href {https://ui.adsabs.harvard.edu/abs/2010ApJ...714..404T} {714, 404}

\bibitem[\protect\citeauthoryear{{Tang}, {MacFadyen}  \& {Haiman}}{{Tang}
  et~al.}{2017}]{Tang17}
{Tang} Y.,  {MacFadyen} A.,   {Haiman} Z.,  2017, \mn@doi [\mnras]
  {10.1093/mnras/stx1130}, \href
  {http://adsabs.harvard.edu/abs/2017MNRAS.469.4258T} {469, 4258}

\bibitem[\protect\citeauthoryear{{Tang}, {Haiman}  \& {MacFadyen}}{{Tang}
  et~al.}{2018}]{tang18}
{Tang} Y.,  {Haiman} Z.,   {MacFadyen} A.,  2018, \mn@doi [\mnras]
  {10.1093/mnras/sty423}, \href
  {https://ui.adsabs.harvard.edu/abs/2018MNRAS.476.2249T} {476, 2249}

\bibitem[\protect\citeauthoryear{{Tazzari} \& {Lodato}}{{Tazzari} \&
  {Lodato}}{2015}]{Tazzari2015}
{Tazzari} M.,  {Lodato} G.,  2015, \mn@doi [\mnras] {10.1093/mnras/stv352},
  \href {https://ui.adsabs.harvard.edu/abs/2015MNRAS.449.1118T} {449, 1118}

\bibitem[\protect\citeauthoryear{{Terashima} \& {Wilson}}{{Terashima} \&
  {Wilson}}{2003}]{Terashima03}
{Terashima} Y.,  {Wilson} A.~S.,  2003, \mn@doi [\apj] {10.1086/345339}, \href
  {https://ui.adsabs.harvard.edu/abs/2003ApJ...583..145T} {583, 145}

\bibitem[\protect\citeauthoryear{{Tiede}, {Zrake}, {MacFadyen}  \&
  {Haiman}}{{Tiede} et~al.}{2020}]{Tiede2020}
{Tiede} C.,  {Zrake} J.,  {MacFadyen} A.,   {Haiman} Z.,  2020, \mn@doi [\apj]
  {10.3847/1538-4357/aba432}, \href
  {https://ui.adsabs.harvard.edu/abs/2020ApJ...900...43T} {900, 43}

\bibitem[\protect\citeauthoryear{{Troja} et~al.,}{{Troja}
  et~al.}{2017}]{Troja17}
{Troja} E.,  et~al., 2017, \mn@doi [\nat] {10.1038/nature24290}, \href
  {https://ui.adsabs.harvard.edu/abs/2017Natur.551...71T} {551, 71}

\bibitem[\protect\citeauthoryear{{Vasudevan} \& {Fabian}}{{Vasudevan} \&
  {Fabian}}{2009}]{vasudevan09}
{Vasudevan} R.~V.,  {Fabian} A.~C.,  2009, \mn@doi [\mnras]
  {10.1111/j.1365-2966.2008.14108.x}, \href
  {http://adsabs.harvard.edu/abs/2009MNRAS.392.1124V} {392, 1124}

\bibitem[\protect\citeauthoryear{Vasudevan, Mushotzky, Winter  \&
  Fabian}{Vasudevan et~al.}{2009}]{vasudevan09b}
Vasudevan R.~V.,  Mushotzky R.~F.,  Winter L.~M.,   Fabian A.~C.,  2009,
  \mn@doi [Monthly Notices of the Royal Astronomical Society]
  {10.1111/j.1365-2966.2009.15371.x}, 399, 1553

\bibitem[\protect\citeauthoryear{{Vaughan}, {Uttley}, {Markowitz},
  {Huppenkothen}, {Middleton}, {Alston}, {Scargle}  \& {Farr}}{{Vaughan}
  et~al.}{2016}]{vaughan16}
{Vaughan} S.,  {Uttley} P.,  {Markowitz} A.~G.,  {Huppenkothen} D.,
  {Middleton} M.~J.,  {Alston} W.~N.,  {Scargle} J.~D.,   {Farr} W.~M.,  2016,
  \mn@doi [\mnras] {10.1093/mnras/stw1412}, \href
  {http://adsabs.harvard.edu/abs/2016MNRAS.461.3145V} {461, 3145}

\bibitem[\protect\citeauthoryear{{Vecchio}}{{Vecchio}}{2004}]{2004PhRvD..70d2001V}
{Vecchio} A.,  2004, \mn@doi [\prd] {10.1103/PhysRevD.70.042001}, \href
  {https://ui.adsabs.harvard.edu/abs/2004PhRvD..70d2001V} {70, 042001}

\bibitem[\protect\citeauthoryear{{Vernstrom}, {Scott}, {Wall}, {Condon},
  {Cotton}, {Kellermann}  \& {Perley}}{{Vernstrom} et~al.}{2016}]{Vernstrom16}
{Vernstrom} T.,  {Scott} D.,  {Wall} J.~V.,  {Condon} J.~J.,  {Cotton} W.~D.,
  {Kellermann} K.~I.,   {Perley} R.~A.,  2016, \mn@doi [\mnras]
  {10.1093/mnras/stw1836}, \href
  {https://ui.adsabs.harvard.edu/abs/2016MNRAS.462.2934V} {462, 2934}

\bibitem[\protect\citeauthoryear{{Volonteri} et~al.,}{{Volonteri}
  et~al.}{2020}]{Volonteri2020}
{Volonteri} M.,  et~al., 2020, \mn@doi [\mnras] {10.1093/mnras/staa2384}, \href
  {https://ui.adsabs.harvard.edu/abs/2020MNRAS.498.2219V} {498, 2219}

\bibitem[\protect\citeauthoryear{{Willingale}, {Pareschi}, {Christensen}  \&
  {den Herder}}{{Willingale} et~al.}{2013}]{willingale13}
{Willingale} R.,  {Pareschi} G.,  {Christensen} F.,   {den Herder} J.-W.,
  2013, arXiv e-prints, \href
  {http://adsabs.harvard.edu/abs/2013arXiv1307.1709W} {}

\bibitem[\protect\citeauthoryear{{Xin}, {Mingarelli}  \& {Hazboun}}{{Xin}
  et~al.}{2021}]{Xin2021}
{Xin} C.,  {Mingarelli} C. M.~F.,   {Hazboun} J.~S.,  2021, \mn@doi [\apj]
  {10.3847/1538-4357/ac01c5}, \href
  {https://ui.adsabs.harvard.edu/abs/2021ApJ...915...97X} {915, 97}

\bibitem[\protect\citeauthoryear{{Yuan}, {Murase}, {Zhang}, {Kimura}  \&
  {M{\'e}sz{\'a}ros}}{{Yuan} et~al.}{2021}]{Yuan2021}
{Yuan} C.,  {Murase} K.,  {Zhang} B.~T.,  {Kimura} S.~S.,   {M{\'e}sz{\'a}ros}
  P.,  2021, arXiv e-prints, \href
  {https://ui.adsabs.harvard.edu/abs/2021arXiv210105788Y} {p. arXiv:2101.05788}

\bibitem[\protect\citeauthoryear{{Zanotti}, {Rezzolla}, {Del Zanna}  \&
  {Palenzuela}}{{Zanotti} et~al.}{2010}]{Zanotti+10}
{Zanotti} O.,  {Rezzolla} L.,  {Del Zanna} L.,   {Palenzuela} C.,  2010,
  \mn@doi [\aap] {10.1051/0004-6361/201014969}, \href
  {https://ui.adsabs.harvard.edu/abs/2010A&A...523A...8Z} {523, A8}

\bibitem[\protect\citeauthoryear{{Zhang}, {Ramos-Ceja}, {Pacaud}  \&
  {Reiprich}}{{Zhang} et~al.}{2020}]{Zhang20}
{Zhang} C.,  {Ramos-Ceja} M.~E.,  {Pacaud} F.,   {Reiprich} T.~H.,  2020,
  \mn@doi [\aap] {10.1051/0004-6361/201937329}, \href
  {https://ui.adsabs.harvard.edu/abs/2020A&A...642A..17Z} {642, A17}

\bibitem[\protect\citeauthoryear{{d'Ascoli}, {Noble}, {Bowen}, {Campanelli},
  {Krolik}  \& {Mewes}}{{d'Ascoli} et~al.}{2018}]{dAscoli18}
{d'Ascoli} S.,  {Noble} S.~C.,  {Bowen} D.~B.,  {Campanelli} M.,  {Krolik}
  J.~H.,   {Mewes} V.,  2018, \mn@doi [\apj] {10.3847/1538-4357/aad8b4}, \href
  {http://adsabs.harvard.edu/abs/2018ApJ...865..140D} {865, 140}

\makeatother
\end{thebibliography}


\bsp	
\label{lastpage}
\end{document}